\documentclass[12pt]{article}

\begin{document}

\title{Theoretical description of 
nucleation in multicomponent system}

\author{Victor Kurasov}

\maketitle

\begin{abstract}
The full theoretical analysis of the kinetics of multicomponent
nucleation is presented. The relief of the free energy
with surface excesses  was analyzed,
the valleys and ridges were described,  their mutual interaction
was studied. The new possibility to change the valley of
nucleation is shown. The possibility to have one common
valley
instead of several neighbor ones which leads to the radical change
in the height of the
effective activation barrier and
to the new value for the nucleation
rate.
\end{abstract}

\section*{Introduction}

Historically the problem of determination of the stationary rate of
nucleation was primary investigated in one-dimensional
approximation \cite{Becker}, \cite{Zeldovich}. The unique variable
characterizing an embryo of a new phase was a number of
 molecules inside the embryo.  Meanwhile, it is evident that the
embryo has at least several characteristics, which have to be taken
into account to give the adequate description of the nucleation
process. That's why it is necessary to study the description of the
nucleation process on the base of several characteristics of the
embryo.

One can not pretend to take into account all characteristics of an
embryo and to give the nucleation description on the base of all
embryo characteristics. The kinetic aspects of the embryo formation
are also far from clear interpretation. Even the
mathematical structure of the theory of
multi-dimension nucleation is far from complete understanding. So,
it is worth to start with the simple cases of multidimension
description.

The simplest and the most evident example of multi-dimension
description is a multicomponent nucleation. It means that the
nucleation in  mixture of vapors is studied. Here kinetic
coefficients are determined extremely clear, the free energy of
the embryo is also rather well known in general features.

The history of investigations of the binary nucleation is very
rich. The number of publications concerning the binary nucleation
is now greater than  devoted to other domains of the
nucleation theory. But already in the case of binary nucleation
there appear many problems to solve. So, it is worth paying
attention namely to multicomponent nucleation.

Until nowadays there is still no universal
self consistent analytic approach which makes use of all previous
theories or directly shows their errors. This task will be the
goal of the present paper.

 At first one has at least to mention approaches, which pretend
 to give original recipes for the stationary nucleation rate.
Certainly, the classical expression for the free energy given by
the standard thermodynamics has to be the starting point of a
theory.
 In
 our analysis we ignore approaches suggesting some artificial
 correction terms or some reconsiderations without  a solid
 thermodynamic base.

 The microscopic corrections to the free
 energy  given by classical
 thermodynamiocs \cite{Hill}, \cite{Schekin}
 are not the subject of our investigation, we consider
 only a  task to describe
 nucleation at the relatively low supersaturations. Even
 this question is out of a true solution. We do not consider
 a
 normalising factor in an
 equilibrium distribution which evidently appear in
 the expression for the nucleation rate. This will be a subject of a
 separate investigation.

The first essential contribution to establish  the binary
nucleation rate was made by H.Reiss \cite{Reiss} who determined
the rate of nucleation on the base of a steepest descent line in
a near critical region. Solution of  a
kinetic equation presented by
Reiss was corrected by Stauffer \cite{Stauffer}.
In the last  paper the
correct formula for the rate of nucleation in the square
approximation of the free energy in the neighborhood of the
critical embryo was given. Earlier the general ideas for the
problem of overcoming the activation barrier in the multicomponent
case were formulated by Langer \cite{Langer} but one can not state
that the publication \cite{Stauffer} is a direct consequence of
\cite{Langer}.
One has to stress that the constant direction of a flow in a
neighborhood of the critical point was simply postulated in
\cite{Stauffer}. This constancy can be proven only with the help of
the boundary conditions
which was done in \cite{Mel2}
where direct solution of the kinetic equation
was presented.

As it became clear after the solution of Trinkaus \cite{Trink}
the problem to determine the nucleation rate
requires to decide  whether the
transition over the barrier really occurs at the critical point ( the
saddle point - here
and later the critical point means the coordinates of the critical
embryo). When there is a strong hierarchy between kinetic
coefficients of absorption
of different components one can see that flow of embryos can
pass aside the critical
region (the region near the critical point),
but over the ridge far from
critical point.

Solution of Stauffer implies the square approximation of
the embryos free energy near the critical point.
Solution of Trinkaus implies
the linear
approximation of the height of special activation barrier.
But as it will be seen later there is no contradiction
between approximations - both are suitable in corresponding
situations.

Further analytical progress is associated with the
appearance of many
variations. Among
them one can outline the refined Stauffer's solution
presented by Berezhkovski and Zitserman \cite{Berej} and conception of
the genuine saddle point proposed by Li et al. \cite{Li}. One has
to stress that these contributions did not radically change the
already known formulas for nucleation rate but slightly corrected
some known results.
In this context
 it is also necessary to mention the publications
of Shi and Seinfeld \cite{shi} and  Wu \cite{Wu}.

Here we do not analyze the theories
 connected with the reconsidered free
energy of the embryos formation taking their history from the
famous publication of Lothe and Pound \cite{Lothe} and
modifications of this approach to the case of multicomponent
nucleation.
Any new expression for the free energy will cause the new
value of the nucleation rate but the mathematical structure
of the derivation of the nucleation rate remains the same.

In the middle of 1990-ies
 the serious set of attempts to analyze the
binary nucleation problems was presented  in \cite{Binexp96},
\cite{BinThe96}.

In the last years one can outline the publications which analyze
the same problems which
have been already mentioned. The problem of
boundary conditions was revised in publication of Wilemski and
Fisenko \cite{Fis}. The authors put the natural boundary conditions
directly at the boundaries of a whole pre-critical region where
these conditions are evident. But then it is necessary to solve
the
kinetic equation in the whole
pre-critical and the near-critical
region which was done in \cite{Fis} only numerically.

The set of papers by Li, Nishioka, Maksimov \cite{Li} is devoted to
give the definition of the generalized saddle point which can be
used both in the
situation where the flow goes over the standard saddle
point and in the
situation of hierarchy where the solution of Trinkaus
\cite{Trink} takes place. This idea is certainly attractive but
as it will be shown in this publication
sometimes the nucleation occurs in a more
complex way and can not be described in terms of the genuine saddle
point even approximately.
Moreover, the point of the Trinkaus' solution depends not only
on the free energy but also on the derivatives along
special directions.

The problem of transition of the binary case to the unary one was
studied in \cite{Wulast} where the full analysis of this problem
was given. Here we are
not interested in this transition because the
embryo with one molecule of a rare component can not be considered
on the base of a standard thermodynamics in an approximation of
a
homogeneous
 liquid which is adopted in this publication.

Here we do not analyze numerous publications which combine the
standard known approach with some artificial additions. Such
combinations are rather typical for publications of Djikaev with
coauthors (see, for example, \cite{dji06a}). In \cite{dji06a} the
values of kinetic coefficients from the first passage time analysis
are
formally
injected in the standard solution presented in \cite{Mel1}
and the final formulas are presented. One has to mention that the
first passage time analysis is based on some unknown
characteristics (for example, the height of activation barrier for
a molecule to penetrate inside the embryo) which can not lead to
concrete results.

All mentioned publications in the binary nucleation do not make any
profit from the topology of the relief of the free energy of
the
embryo. This task was solved in \cite{Mel1} where the structure of
relief of the free energy in the capillary approximation was studied.
It was shown that the relief of the free energy can be
characterized as the straight channels, ridges and saddle
points. In this publication the results of \cite{Mel1} will be
widely used.

The formulation of the capillary approximation faces the difficulty
known as the Renninger-Wilemski's paradox \cite{Renn}, \cite{Wil}.
Because of publications by Oxtoby and Kashiev \cite{OxtKash} the
thermodynamic background of the surface excesses is completely
studied.
To overcome this difficulty one has to write the Gibbs' absorption
equation and to introduce surface excesses of components of the
surface of tension. This leads to the difference of concentration
in the surface layer and in the bulk of the
embryo which was noted in
\cite{Mel1}. But  there  further conclusions for
kinetics of the process have not being made.

The structure of the free energy relief with surface excesses was
investigated in \cite{himphys} but only in thermodynamic aspects of
the problem. The kinetic features have not been considered in
\cite{himphys}.

In \cite{Mel2} the kinetic equation was solved  in the
neighborhood of the critical embryo. The progress achieved
in \cite{Mel2} was the appropriate formulation and account of
boundary conditions. Certainly earlier the boundary conditions
were mentioned in \cite{Trink} but they were put in the infinitely
far points where the structure of a free energy can not be
seen in all details. Namely the necessity to conserve
the boundary conditions at the low boundary of a near-critical
region determines the conservation of the square form of the free
energy in transformation presented
in \cite{Mel2}.

Having summarized the development of theoretical
investigations in the binary nucleation one can state that
despite the essential progress in this field there are
still many problems to consider.

It is rather natural to construct the global picture of the
nucleation
including the case of the hierarchy between kinetic coefficients,
surface excesses, etc. The unification of the free energy
topological features with the already mentioned approaches is the
main goal of this publication. This approach leads to many rather
essential features of nucleation presented below. Moreover, some
striking features changing the rate of nucleation in the order of
magnitude will appear.

The structure of this paper is the following
\begin{itemize}
\item
In the first part the main ideas of the capillary approximation are
formulated and the free energy is constructed. Here
the  surface excesses are
taken into account. The variables providing the simple form of the
free energy are shown and their connection with the numbers of
molecules in the embryo is established.
\item
The second part is devoted to the description of the near-critical
region. It is shown
that this region
has the form similar to the case of the absence of the
surface excesses. Here the hierarchy of evolution will be shown.
\item
The third part is devoted to the analysis of the Reiss' solution
and the Stauffer's one. The plausible way to see the Reiss' formula
will be shown. The moderate value of the difference
between the Reiss' and the Stauffer's solution is justified.
This is important for the possible ignorance  of the slow
or rapid
variables of correcting order.
\item
The forth section analyzes the jump of the
embryos from one channel to
another one.
The situation of the near-equilibrium falling transition
is considered here.
The solution is found also in the discrete model.
\item
The fifth  section
considers the conception of the common valley.
The equilibrium common valley transition is analyzed.  It
will be shown  the new height of activation barrier.
 This value seriously differs from all known results.
\item
The sixth section analyzes the general picture picture of
the nucleation rate formation. The case of the equilibrium
saturation of the destination valley is studied.
\item
All results are summarized in the conclusion.
\end{itemize}

\section{Thermodynamic basis}

\subsection{Capillary approximation}

The main object involved in determination of the nucleation rate is
the free energy of the isolated embryo. To give the description of
the embryo one has to fix the variables of the state of the
embryo. Assuming the thermal equilibrium of the embryo one can
describe the embryo
only by the numbers of molecules $\nu_i$ inside the
embryo. These variables are extracted by following properties
\begin{itemize}
\item
In elementary acts of evolution $\nu_i$ are changed separately. The
step of change is one unit.
\item
Although the free energy even in capillary approximation is not
diagonal the form of expression for the free energy is relatively
simple.
\end{itemize}

In the capillary approximation the energy $F$ ordinary taken in the
thermal units is the sum of the bulk part $B$ and the surface part
$\Omega$.
$$
F =  - B + \Omega
$$
The ordinary expressions for $B$ and $\Omega$ are following
$$
B = \sum_i
\nu_i \mu_i
$$
$$
\Omega = \gamma S
$$
Here the sum is taken over all components of the embryo, $\mu_i$
are the differences of the chemical potentials counted from the
equilibrium values (with a negative sign),
 $S$ is the square of the surface of tension,
$\gamma$ is the renormalized surface tension.

The difference between the precise value of the free energy and the
value in the capillary approximation referred as "correction terms"
($c.t.$) is supposed to be relatively small in comparison with
$B+\Omega$. This situation takes place when the number of molecules
$$\nu_{tot} = \sum_i \nu_i $$
inside the embryo is very (strictly
speaking infinitely) big
\begin{equation} \label{totbig}
\nu_{tot} \gg   1
\end{equation}
The inverse number of molecules (i.e. $\nu_{tot}^{-1}$)
 will be the small parameter of the
theory. So,
$$
F= - B+\Omega + c.t.
$$
where $c.t.$ indicates correction terms with a property
$$
|c.t.| \ll |F|
$$
Ordinary the decomposition of correction terms on inverse radius
$r^{-1}$
of the embryo converges and $F$ has the form
$$
F= - B+\Omega + \sum_{k=-1}^{\infty} c_k r^{-k} + c_0 \ln r
$$
Here $c_k$ are the coefficients. One
can also consider the last decomposition
as an asymptotic decomposition. We shall accept the validity of
this decomposition.

Ordinary this decomposition is taken with a finite number
of terms
\begin{equation}\label{finite}
F= - B+\Omega + \sum_{k=-1}^{1 \div 2} c_k r^{-k} + c_0 \ln r
\end{equation}

From the last decomposition it follows
$$
 | d \ c.t.| \ll | d F |
$$
The last inequality is important in the justification of
the linearization of the free energy.

While speaking about the capillary approximation one has to imply a
whole set of assumptions
beside the pure thermodynamic consideration.
There are several ordinary used
approximations included into the capillary approximation. These
approximations are the following
\begin{itemize}
\item
The
 surface tension is attributed to the dividing surface
calculated on the base of the volume separation, i.e.
$$
S = (\sum_i v_i \nu_i)^{2/3}
$$
where $v_i$ are the volumes in a liquid phase. The formal factor $4
\pi / (4 \pi / 3)^{2/3}$ is ordinary included into the effective
surface tension.
\item
Values $v_i$, $\gamma$ are taken from the  case of a bulk liquid.
\item
To give expressions for $\mu_i$ one has to use some model. The most
widely used model is the model of
a liquid solution. The validity
of this model  requires
$$\nu_i
\gg 1 $$
 for every component. Certainly one can use other
models and ignore  these limitations. When $\nu_j =1$ for some
component, one can consider this component as  a heterogeneous
center. That's why the extension  of the approximation of
the regular solution up to $\nu_i=1$ in \cite{Wulast} causes
questions.

In this paper  we shall use the model of solution.

To give a
formula for chemical potential one can define a supersaturation as
$$
\zeta_i = \frac{n_i}{n_{i\infty}}
$$
where $n_i$ is the molecular number density in the existing vapor
and $n_{i\infty}$ is the molecular number density of the vapor
saturated over the pure bulk liquid of component $i$
with a flat surface. Then
$$
\mu_i = \ln(\zeta_i) - \ln \xi_i - \ln f_i(\{\xi\} )
$$
Here it is supposed that the vapor is an ideal gas which gives the
value for the first term in the r.h.s. as
$  \ln(\zeta_i) $. Ordinary it is
assumed that the concentrations $\xi_i$ form a set $\{\xi \}$ of
concentrations and the coefficients of activity $f_i$ can depend on
the whole set of activities.

For approximation of ideal solution all coefficients
$$f_i=1$$
To know $f_i$ one has to construct some model of
solution or to use some experimental data.

\item
A special question concerns the definition of concentration.
Ordinary the concentration  is determined as
\begin{equation} \label{c}
\xi_i = \frac{\nu_i }{\sum_j \nu_j}
\end{equation}
Sometimes this definition is also included into the ordinary
auxiliary approximations of the capillary approach.

This question is directly linked with the Wilemski-Renninger's
paradox \cite{Renn}.

\end{itemize}

All assumptions made above are necessary for formula for the free
energy in capillary approximation.

The presented formula for the free energy is rather transparent,
 but it faces the difficulty known as the Wilemski-Renninger's
paradox. The difficulty is the following:
\begin{itemize}
\item
It is known that in the critical embryo the Kelvin's relation
$$
\frac{\mu_i}{v_i} = invariant
$$
has to be observed. This follows from the general thermodynamics
and from the sense of chemical potentials.

The last relation gives an
equation for the concentration in the critical embryo.
\item
One can come to the same equation on concentration also by direct
differentiation of expression for the free energy. For simplicity
assume that $v_i$ do not depend on concentration. This gives
$$
\frac{\partial F}{\partial \xi_i } = 0
$$
i.e.
$$
0
=
- \mu_i  - \sum_j \frac{\partial \mu_j}{\partial \nu_i} \nu_j +
\frac{2}{3} \gamma (\sum_j v_j \nu_j )^{-1/3} v_i + S \frac{\partial
\gamma}{\partial \nu_i}
$$
Here it is supposed that $v_i$ do not depend on
concentration.

One has to recall
 that the coefficients of activity $f_i$ satisfy the Gibbs-Duhem's
equations
$$
\sum_i \xi_i d \mu_i = 0
$$
which put a restriction on the coefficients of activity.
$$
\sum_i \xi_i d \ln f_i = 0
$$
Moreover the Gibbs-Duhem equation can be written as
$$
\sum_j \frac{\partial \mu_j}{\partial \nu_i} \nu_j = 0
$$
Then the differentiation becomes very simple and leads to
$$
\frac{\partial F}{\partial \xi_i } = 0 =
- \mu_i   +
\frac{2}{3} \gamma (\sum_j v_j \nu_j )^{-1/3} v_i  +
S \frac{\partial
\gamma}{\partial \nu_i}
$$
Then one can come to the widely known Kelvin's equation only if the
derivative $\partial \gamma / \partial \xi_i$ is zero. So, the
formal recipe is to forbid the differentiation of the surface
tension on concentration. Since the last equation comes from the
foundations of thermodynamics it means that something is irrelevant
in the previous formula for the free energy.

In the case when $v_j$ depend on concentration we have
$$
\frac{\partial F}{\partial \xi_i } = 0 =
- \mu_i  - \sum_j \frac{\partial \mu_j}{\partial \nu_i} \nu_j +
\frac{2}{3} \gamma (\sum_j v_j \nu_j )^{-1/3} [v_i + \sum_j
\frac{\partial v_j}{\partial \nu_i } \nu_j ] + S \frac{\partial
\gamma}{\partial \nu_i}
$$
But the Gibbs-Duhem equation has to be here the
following one
$$
- \sum_j \frac{\partial \mu_j}{\partial \nu_i} \nu_j +
\frac{2}{3} \gamma (\sum_j v_j \nu_j )^{-1/3}  \sum_j
\frac{\partial v_j}{\partial \nu_i } \nu_j  = 0
$$
which leads to the same conclusions.

As the result there appeared a formal recipe not to differentiate
the surface tension. At first it was the artificial recipe but
later
 the justification of this
recipe was given on the base of the Gibbs dividing surfaces
formalism.

To resolve  this difficulty  one has to add to the free
energy some new contributions connected with the surface
excesses.
It will be done later.

\end{itemize}

Now we return to consideration of the properties of $F$.

The leading idea here is the extraction of
the mentioned small parameters $\nu_i^{-1}$.
Recall that conditions are rather far from the second order
phase transition. If we accept that the
surface layer has a finite thickness $d$, then in the limit $r
\rightarrow \infty$ (where $r$ is the radius of the embryo)
one can see that correction terms ($c.t.$) are really relatively
small in the following sense
$$
| B  | \gg  | c.t. | \ \ \ \ \ | \Omega |  \gg | c.t. |
$$
$$
| \frac{\partial c.t.}{\partial \nu_i}  | \ll |
\frac{\partial B}{\partial \nu_i} |
\ \ \ \ \ \
| \frac{\partial c.t.}{\partial \nu_i}  | \ll |
\frac{\partial \Omega}{\partial \nu_i} |
$$
$$ |
\frac{\partial^2 c.t.}{\partial \nu_i^2}  | \ll |
\frac{\partial^2 \Omega}{\partial \nu_i^2} |
$$
These inequalities are valid for absolute values.

These  inequalities is a new result and
they
 will be widely used below. Their validity
 can be proven analytically.

\subsection{The form of the free energy}

To see the structure of the free energy one can introduce the
extensive variable
$$
V = \sum_i v_i \nu_i
$$
Certainly $V$ is a volume of the embryo. Then
$$
F = \gamma V^{2/3} - b(\xi)V
$$
with the generalized chemical potential
$$
b = \frac{\sum_j \mu_j \xi_j}{\sum_j v_j \xi_j}
$$
The generalized chemical potential allows
an interpretation
$$
b = \frac{< \mu >}{< v >}
$$
as the ratio of the mean chemical potential excess and the mean
volume per one molecule in the embryo.

One can take also as an extensive variable the total number of
molecules inside the embryo
$$
\nu_{tot} = \sum_i  \nu_i
$$
Then the free energy can be written in a following way
$$
F = \gamma (\sum_i v_i \xi_i)^{2/3} \nu_{tot}^{2/3} - \nu_{tot}
\hat{b}(\xi)
$$
where
$$
\hat{b}(\xi) = \sum_i \mu_i \xi_i
$$
and the renormalized surface tension
$$\hat{\gamma} = \gamma (\sum_i v_i
\xi_i)^{2/3}
$$
appears.

One has also to mention the possibility to take as external
variable the surface energy in the power $3/2$, i.e.
$$
\varsigma = \gamma^{3/2} \sum_i v_i \nu_i
$$
used in \cite{Mel1}. Then the free energy has the form
$$
F = - b_p(\xi) \varsigma + \varsigma^{2/3}
$$
where the generalized chemical potential is
$$
b_p = \frac{\sum_j \mu_j \xi_j}{\sum_i \gamma^{3/2} v_i \xi_i } =
\frac{b}{ \gamma^{3/2}  }
$$

The third possibility used in \cite{Mel1} is the most preferable
because here the free energy has the most simple form and in the
"surface" term the factor depending on concentration is absent.

The question to discuss is what we shall take as an
extensive variable - the variable proportional to the
volume or the value proportional to the number of
molecules? Thermodynamics does not give an answer because
asymptotically these values are proportional.

But the problem to take into account the Renninger-Wilemski's
paradox remains here our of attention. To overcome this difficulty
one has to include into description the surface excesses of
components. To take these excesses one has to choose
  the
surface
accurately. The most preferable
choice is to choose as the surface the surface of
tension
because the surface
tension can be attributed
 to this surface without corrections.
At this surface
 all components have the surface excesses $\psi_i$ but
the surface tension can be attributed namely to this surface. In
the first (rough)
approximation these values are proportional to the square
$S$ of the surface of tension
$$
\psi_i = \varrho_i S
$$

Parameters $\varrho_i$ are supposed to be independent on $S$ and
have to be given by the theory of a liquid state.

The square of the surface of tension can be
approximately calculated as
$$
S = (\sum_i v_i (\nu_i - \psi_i ))^{2/3}
$$
Certainly, there exists a difference between
a surface of tension and the surface covering the volume of the
embryo. But since the sense has only $S\gamma$ one can attribute
this difference to the value of $\gamma$.

Here we omit the constant factor having included it into the
surface tension $\gamma$.
Hence,
$$
\psi_i = \varrho_i (\sum_i v_i (\nu_i - \psi_i ))^{2/3}
$$
The last relation is not a formula for $\psi_i$
but an equation. It
can be solved by iterations. These iterations
are based on a small parameter $\psi_i / \nu_i $. The smallness of
these parameters at $\nu_i \rightarrow \infty$ is evident.
The first approximation
$$
\psi_i = \varrho_i (\sum_i v_i \nu_i)^{2/3}
$$
is already suitable as a leading term under the conditions
(\ref{totbig}). The second iteration
$$
\psi_i = \varrho_i (\sum_i v_i (\nu_i - \varrho_i (\sum_j v_j \nu_j)^{2/3} ))^{2/3}
$$
will refine the solution. The complexity of dependence of $\psi_i$
on $\nu_i$ is the certain difficulty.

The value of concentration $\xi_i$ has now to be redefined as
$$
\xi_i = \frac{\nu_i - \varrho_i S }{\sum_j (\nu_j - \varrho_j S )}
$$

As an extensive
variable it is natural to choose the straight analog
of $\varsigma$, namely
$$
\varsigma = ( S \gamma )^{3/2} =
\sum_i v_i (\nu_i - \psi_i )\gamma^{3/2}
$$
But this choice does not lead to the "true" form of the free but to
$$
F =  - \sum_j \lambda_j \mu_j - \sum_j \varrho_j \mu_j
\frac{\varsigma^{2/3}}{\gamma(\xi)} +\varsigma^{2/3}
$$
with
$$
\lambda_i =
\nu_i - \psi_i
$$
Here the dependence $\gamma$ on $ \{ \xi \} $ is the source of
difficulties. Certainly,
$$
\frac{\lambda_i}{\lambda_j} = \frac{\xi_i}{\xi_j}
$$

One can introduce another set of variables.
Now instead of $\varsigma$ one has to choose the extensive variable
$$
\kappa = S^{3/2} ( \gamma  - \sum_i \varrho_i \mu_i )^{3/2}
$$
In these variables the free energy $F$ has the form
\begin{equation} \label{form}
F =  - \kappa b_g (\xi) + \kappa^{2/3}
\end{equation}
with the generalized chemical potential
$$
b_g = \frac{\sum_i \lambda_i \mu_i}{\kappa}
$$
or
$$
b_g = \sum_i \xi_i \mu_i \frac{\sum_j \lambda_j}{\kappa}
$$
One has to show that $b_g$ does not depend on $\kappa$. To
fulfill this derivation  one
can come to
$$
b_g = \sum_i \xi_i \mu_i \frac{\sum_j \lambda_j}{S^{3/2} (\gamma -
\sum_k \varrho_k \mu_k)^{3/2} }
$$
or
$$
b_g = \sum_i \xi_i \mu_i \frac{\sum_j \lambda_j}{ (\gamma - \sum_k
\varrho_k \mu_k)^{3/2} \sum_l v_l \lambda_l }
$$
It can be also presented as
\begin{equation}\label{13}
b_g = \sum_i \xi_i \mu_i \frac{1}{ (\gamma - \sum_k \varrho_k
\mu_k)^{3/2} \sum_l v_l \xi_l }
\end{equation}
The last relation evidently shows that $b_g$ is really a function
of $\xi$. The dependence on $\kappa$ is absent.

One can use  expression (\ref{13}) to clarify the
Renninger-Wilemski's paradox.
According to the Gibbs' absorption relation
$$
d\gamma =  d \sum_j \varrho_i \mu_i
$$
the derivative of the surface tension on concentration is cancelled
by the corresponding derivatives of $\varrho_i$ on $\xi$. So,
if we write $b_g$ without surface excesses as
$$
b_g = \sum_i \xi_i \frac{1}{\gamma^{3/2} \sum_j v_j \xi_j}
$$
we have to forbid the differentiation of $\gamma$ on
concentration.
Now
the Renninger-Wilemski's paradox is explained. It is necessary to
stress that the reason is not the formal
Gibbs' absorption equation,
but the difference of concentrations in the bulk solution from the
integral values.

Although the the new variables ensure the simple form of the free
energy their connection with "initial" variables $\nu_i$ is rather
complex. One has to see how on the base $\kappa$, $\xi_i$ it is
possible to reconstruct $\nu_i$.
The procedure is the following:
\begin{itemize}
\item
 On the base of $\xi_i$ we know
$\mu_i$, then we get $\gamma - \sum_i \varrho_i \mu_i$. \item
This gives
a value of
$$S= \kappa^{2/3} / (\gamma - \sum_i \varrho_i \mu_i)$$
\item
 On the base
of $S$ having presented $S$ as
$$S= \sum_i v_i \lambda_i =
\sum_i v_i \xi_i  \sum_j \lambda_j$$
 we get
$\sum_i \lambda_i$.\item
 Since $\lambda_i = \xi_i \sum_j \lambda_j$ we
get all $\lambda_i$. \item
Then
$$\nu_i = \lambda_i + \varrho_i (\xi) S$$
and we know all $\nu_i$.
\end{itemize}

 The inverse transformation can not be made
by explicit formulas, the problem to find\footnote{When the index
is absent it means that the whole set is considered.} $\varrho$ on
the base of $\nu$ has been considered above. When $\varrho$ is
found then $\lambda$ is known. This gives $\xi$ and $\kappa$.

The main new facts found here are the following:
\begin{itemize}
\item
The variables giving the simple expression for the free energy
with surface excesses are
found.
\item
The recipe to get the initial variables on the base of the new ones
is given
\end{itemize}

\subsection{The structure of the free energy relief}

The functional form (\ref{form}) has some consequences analogous to
those considered in\footnote{In \cite{Mel1} the free energy without
surface excesses was considered. } \cite{Mel1}.
But now this form takes into account the surface
excesses of an embryo. Here the form
(\ref{form}) ensures the following properties of the free energy of
an embryo
\begin{itemize}
\item
One can see the channels of nucleation defined by equations
$$
\frac{\partial b_g}{\partial\xi_i} = 0
$$
$$
\frac{\partial^2 b_g}{\partial\xi_i^2} < 0
$$
Along these channels the equilibrium density of distribution has a
maximum (but the real distribution coincides with the equilibrium
one only in the part of the pre-critical region)
\item
Because of the Gibbs-Duhem's equation the variables $\xi_i$ in
differentiating of $b_g$ are separated. This leads to the
approximately zero
value of the cross derivatives $\partial^2 b_g /
\partial \xi_i \partial \xi_j$.
\item
One can see the separation lines of nucleation defined by equations
$$
\frac{\partial b_g}{\partial \xi_i} = 0
$$
$$
\frac{\partial^2 b_g}{\partial\xi_i^2} > 0
$$
Along the separation lines the equilibrium density of distribution has
a minimum.
\item
In one channel there is only one saddle point. Certainly, this
takes place only in the capillary approximation. This saddle point
has a coordinate $\kappa_c$ determined from the following equation
$$
\kappa_c = (\frac{2}{3b_g(\xi_c)})^3
$$
Here $\xi_c$ is the coordinate of the channel.
\item
The amplitude value of the free energy $F_c$ in the channel is
given by the formula
$$
F_c = \frac{1}{3} \kappa_c^{2/3}
$$
Here one can see the Gibbs' equation and now it is clear that
namely $\kappa^{2/3}$ is the true surface energy, but not $\gamma
S$ as it seems from the first point of view. One has to attribute
to the surface energy all energy like contributions with the space
dimension $2$ (or $2/3$ in relative units).
\item
All channels are independent - the embryos starting from the origin
of coordinates will use only one separate channel to go to the
supercritical region where they begin to grow irreversibly. The
nucleation flow will mainly go through the channel with minimal
$\kappa_c$ or maximal $b_g$. This remark concerns the case where
there is no strong hierarchy between kinetic coefficients of
absorption.
\end{itemize}

One can see that the picture of nucleation is rather simple, but
this simplicity was observed for the free energy with the surface
excesses for the first time here. This is the new result of this
section.

\subsection{The  form of the near-critical region}

As it has been mentioned at the beginning the set of natural
variables is $\nu_i$. The elementary kinetic act of absorption
leads to the change
$$\nu_i \rightarrow \nu_i \pm 1 $$
 So,
it is necessary to establish connection between $\kappa, \xi$ and
$\nu$ at least approximately.

 Denote by the subscript $o$ the values when all surface  excesses
 are zero. Then the theory is very simple and one can get the
 connection between $\kappa_o, \xi_o$ and $\nu_o$ in a very
 transparent
 manner.
 From $\nu_0$ to $\kappa_0, \xi_0$ one can get by
 $$
 \kappa^{2/3}_0 = \gamma \sum_i  v_i \nu_{i0}
$$
$$
\xi_{i0} = \nu_{i0} / \sum_j \nu_{j0}
$$
Inverse transformation is given by the chain formulated above. So,
it is quite easy to write
the kinetic equation for the case of the
absence of excesses.

The above consideration  shows the role of the case with zero
excesses. Hence, this case will be the base to construct the
description in the general case.

Return now to the general case.

One can define the near-critical region as the region where
$$
| F - F_c | \leq 1
$$
This is quite analogous to the one component case.
But here we consider the near-critical region associated
with the given channel. Then it is necessary that this
region  has to be closer to this point than the
separation lines.

One can define the positive size $\Delta \kappa$ of the
near-critical region along the channel of nucleation as
$$
F(\kappa_c \pm \Delta \kappa, \xi_c ) = F_c - 1
$$
Here $\xi_c$ is the coordinate of some channel. Certainly, we get
two values $\Delta_1 \kappa$ and $\Delta_2 \kappa$
 corresponding to the positive and to
the negative shift. In the square approximation
of the free energy
\begin{equation} \label{yy}
\Delta_1 \kappa = \Delta_2 \kappa = 3 \kappa^{2/3}
\end{equation}
When $\nu_i \gg 1$ for all $i$ the square approximation is
rather accurate.

Analogously one can define the characteristic sizes $\Delta \xi_i$
according to relation
$$
F(\kappa_c, \xi_c \pm \Delta \xi_i ) = F_c + 1
$$
Certainly, we get two values $\Delta_1 \xi_i$ and $\Delta_2 \xi_i$
 corresponding to the positive and to
the negative shift. In the square approximation
\begin{equation} \label{yy1}
\Delta_1 \xi_i
= \Delta_2 \xi_i =
|\frac{\partial^2 b_g}{2 \partial \xi_i^2}|^{-1/2} \kappa^{-1/2}
\end{equation}
When $\nu_i \gg 1$ for all $i$ and there is no singular behavior of
generalized chemical potential then the square approximation is
valid.

Certainly, it is necessary that the channels have to  be
separated, i.e. the height of the separation line has to be
several thermal units higher than the height of the
channels. This has to take place at $\kappa$ near the
critical value.

We define the reduced near-critical region as the region where
$|\kappa-\kappa_c| \leq  \Delta \kappa$,
$|\xi_i-\xi_{i c}| \leq  \Delta \xi_i$. This definition differs
from the ordinary definition of the near critical region as
extracted by condition $|F-F_c| \leq 1$.

In the multi-dimensional case there exists long tails near
lines $F=F_c$. To illustrate it one can use the square
approximation, then the curves $F=F_c+1$ and $F=F_c - 1$ are
hyperbolic ones with common asymptotics which are straight lines.

We shall define the tails as the regions corresponding to $|F-F_c|
\leq 1$ and
$|\xi_i - \xi_{ic}| > \Delta \xi_i $, $|\kappa - \kappa_{c}| >
\Delta \kappa $.

Actually, the following statements can be proven analytically:
\begin{itemize}
\item
One can show that the tails do not play any essential role in
formation on the total nucleation flow.
\item
Then it is possible to
reduce the near-critical region up to the following domain
$$|\xi_i - \xi_{ic}| \leq \Delta \xi_i $$
$$|\kappa - \kappa_{c}| \leq \Delta \kappa $$
\end{itemize}
Here and later we shall imagine the reduced near-critical region
 speaking about the near-critical region.

Now one can see that the relative sizes of the near-critical region
are small
 $$|\xi_i - \xi_{ic}| \ll \xi_i $$
$$|\kappa - \kappa_{c}| \ll  \kappa $$

Ordinary this smallness is implied when the kinetic coefficient of
absorption is supposed to be a constant value. Here
this smallness
 will help  to prove the following main result of this
section:
\begin{itemize}
\item
{\bf In the
near-critical region
the function $F - F_c $ as a function of variables $\kappa -
\kappa_{c}$, $\xi_i -
\xi_{ic}$ for every $i$ has practically the same behavior as the
function $F_o - F_{co} $ as a function of variables $\kappa_o -
\kappa_{co}$, $\xi_{io} -
\xi_{ico}$. At least the relative difference is small: }

$$
 \frac{| (F(\nu_i - \nu_{ic}) - F_c )  - ( F_o ( \nu_{io} - \nu_{ico} ) - F_{co} )
 |}{( F_o ( \nu_{io} - \nu_{ico} ) - F_{co} )} \ll 1
 $$

The explanation and the idea of the proof is rather simple.
Really, the correction terms to which the excesses belong begin to
be essential only when the surface term cancels the bulk term. But
as it clear from the sequential differentiation this can take place
only in the first derivative. Starting from the second derivative
the contribution from the bulk term is zero and this compensation
can not take place.
 This effect is taken into account by the shift of $\nu_{ic}
 $ instead of $\nu_{ic0}$. So, here the influence of
 correction termms is negligible.

\end{itemize}

The last result allows to write the kinetic equation in $\nu_i$
variables taking into account the surface excesses by a simple
shift. This takes place only in the near-critical region. This
result is new.

\subsection{The place of the Renninger-Wilemski's effect}

The "paradox" of Wilemski and Renninger
occupies so important place in
the multicomponent nucleation that
from the first point of view it seems that this is the real
effect taking place in the leading term of capillary approximation.
Below it is shown that this effect has an order of correction.
To see this effect
one can redefine $\kappa$ as $S^{3/2}$ and forget
about excesses.

Really, from equation
$$
\frac{\partial F }{ \partial \xi }|_{\kappa = fixed}
=
\frac{\partial \gamma  }{ \partial \xi } \kappa^{2/3} +
\frac{\partial b_g}{\partial \xi}
\kappa \gamma^{3/2}
$$
it is seen that the first term $\frac{\partial \gamma }{ \partial
\xi } \kappa^{2/3}$ with the derivative $\frac{\partial \gamma }{
\partial \xi }$ has a correction order $\kappa^{2/3}$
$$
\frac{\partial \gamma }{ \partial
\xi } \kappa^{2/3}
\sim
\kappa^{2/3}$$
in comparison
with the second term $\frac{\partial b_g}{\xi}
\kappa \gamma^{3/2}
$ having the order $\kappa$

$$\frac{\partial b_g}{\xi}
\kappa \gamma^{3/2}
\sim \kappa$$

We extract this result which is explicitly outlined for the first
time here because of its importance for the reconstruction of the
logical self-consistency of thermodynamics. Only the correcting
order of the term with the derivative of the surface tension allows
to ignore it in the main order and to return the leading role of
the ordinary capillary approximation.

Since the formal recipe to resolve
the Renninger-Wilemski's paradox is to
forbid the differentiation of $\gamma$ on concentration
then the
equation on concentration will be different. It would cause the
impression that there is a shift in a leading term. The correct
answer is that this result causes the shift in $F_c$ which has a
correction order as it follows from the last equation.

The necessity to develop the theory with surface excesses
is evident because the surface excesses  will
essentially shift the position of the near-critical region. The
shift is many times greater than the size of the near-critical
region. The shift has the order  $\kappa$ (because there is
another equation on concentration - the derivative of $\gamma$ on
$\xi$ is cancelled) while the size of the near-critical region has
the order $\kappa^{1/2}$.

One can treat the surface tension as a coefficient in the
first correction term proportional to the surface of the
embryo. The coefficients at $\kappa^{1/3},
\ln \kappa,\kappa^{-1/3}$, etc. depend on intensive
variables (concentrations is one example). Their derivatives
will be cancelled by derivatives of corresponding excesses.
The structure will resemble the Renninger-Wilemski's
paradox. But here the dimension of "surface" will be  $\kappa^{1/3},
\ln \kappa,\kappa^{-1/3}$, etc. This effect will be called
as "generalized cancellation of derivatives on intensive
variables".

One to note that the same procedure can be effectively applied for
all other correction terms. Rigorously speaking to determine the
form of the near-critical region one has to take the expression
for $F$ with correction terms up to the order which causes the
shift of position of the near-critical region. Now it is clear
that the effect of all correction terms will be quite similar to
the already described one.

\section{Channels and separation lines }

\subsection{Similarity of the near-critical relief}

Although the Renninger-Wilemski's effect has a correction order it is
worth taking it into account. The main reason is the following:
\begin{itemize}
\item
The relative sizes of the near-critical region is very small.
Really, from (\ref{yy}) it follows that
$$
\frac{\Delta \kappa}{\kappa_c} \sim \kappa_c^{-1/3}
\ll 1
$$
and the relative size in $\kappa$-scale is small. From (\ref{yy1})
it follows that
$$
\Delta \xi_i / \xi_i \sim \kappa^{-1/2}
$$
and the relative size in the $\xi_i$ scale is small also. Then it
is clear that the relative size in $\nu_i$ scale will be
$$
\nu_i / \nu_{ic} \sim \nu_i^{-1/3}
$$
Namely these estimates allow to put in the near-critical region the
kinetic coefficient $W^+_i$ of absorption of the molecule of $i$-th
component to the constant value $W^+_{ic}$ corresponding to the
critical embryo
$$
W^+_i \approx W^+_{ic}
$$
\end{itemize}

So, the relatively small error in the determination of the
coordinates $\nu_i$ can remove embryo out of the near-critical
region which makes the consideration of kinetic equation without
surface excesses in the near-critical region useless.

Beside this one has to take into account that the elementary
transitions are written in the $\nu$-scale
$$
\nu_i \rightarrow \nu_i \pm1
$$
and the free energy is written explicitly (with the surface
excesses
account) in variables $\kappa$, $\xi_i$. So, it is necessary to
have the a very precise transformation between $\nu_i$ and $\kappa,
\xi_i$. This forms the problem.

Although the transformation from $\kappa, \xi_i$ to $\nu_i$ exists
it is very complex. The inverse transformation has not been
found explicitly. So, it
is necessary to establish the approximate connection. The following
statement establishes this connection
\begin{itemize}
\item
The function $F-F_c$ as a function of $\nu_i - \nu_{ic}$
approximately coincides in the near-critical region
with the behavior of $F_0 - F_{0c}$ as a
function of $\nu_i - \nu_{ic0}$:
$$
F_0 (\nu_{ic0} + y_i) - F_{c0} \approx F (\nu_{ic} + y_i) - F_{c}
$$
\end{itemize}

This property can be called as the approximate similarity of the
free energy relief.

Here this fact is established for
$\{ \nu_i \}$ variables while earlier the same conclusion was made
for $\kappa, \{ \xi \}$ variables.

The idea of the proof of this property is based on the simple
remark that the terms produced by the surface excesses can be
important only when the terms produced by the bulk and surface
contributions are cancelled. In the near-critical region this
occurs only in the first derivative
over $\nu_i$ at the critical embryo. Cancellation
in high derivatives is impossible\footnote{Since the high
derivatives of the bulk contribution are zero.}.
Here we use the form $F=\sum_i \mu_i \nu_i - \gamma S$ and
differentiate it over $\nu_i$. This ensures the
similarity of relief.

Now one can propose the following sequence of actions
\begin{itemize}
\item
 at first
one has to solve equations for the characteristics of the critical
embryo
\item
 then one has to solve the kinetic equation without excesses but
in shifted coordinates.
\end{itemize}

Certainly, the similarity of relief takes place both in $\nu_I$
coordinates and in $\kappa, \xi_i$ coordinates.

Analogously one can one can prove the small relative role
of microscopic corrections in the value of $dF_c/d\zeta_i$
which is used in construction of the global evolution of
the phase transition. Here the formula (\ref{finite})
has to be  used and it has to be taken into account that
the coefficients
in this formula are constant.

\subsection{The form of pre-critical region }

One can see the following important property:
\begin{itemize}
\item
{ \bf The critical embryo can not have $\xi_i = 0$ }
\end{itemize}
It can be seen from the explicit form of $\mu_i$
$$
\mu_i = \ln \zeta_i + \ln \xi_i + \ln f_i(\xi)
$$
Recall that
here $\zeta_i$ is the supersaturation of $i$-th component defined
as
$$
\zeta_i =
\frac{n_i}{n_{ii}}
$$
$n_i$ is the molecular number density
of vapor of $i$-th component, $n_{ii}$
is the molecular number of the pure saturated vapor of $i$-th
component. The second term is caused by the standard entropy of
mixing, the third term characterizes the deviation of mixture from
the ideal solution, here $f_i$ is the phenomenological coefficient
of activity.

Then one can see that at $\xi_i \rightarrow 1$ the situation of
dilute solution takes place. Then the Henry's law states that the
situation is close to the ideal solution, then $f_i =1 $ and there
are no correction terms. Then one can see that
$$
\frac{db_g}{d\xi_i}|_{\xi_i = 1} = \infty
$$
and the condensation into the pure component is forbidden.
Analogously
$$
\frac{db_g}{d\xi_i}|_{\xi_i = 0} =  - \infty
$$

Earlier the analogous estimates were formulated in \cite{Mel1} for
$\xi_{i0}$. Then from (\ref{yy}) and (\ref{yy1}) it follows that
the widths $\Delta \nu_i$ along $\nu_i$ satisfy
$$
\Delta \nu_i \gg 1
$$
for all $i$. These estimates ensure the possibility of continuous
description of evolution in the kinetic equation.

In the absence of the strong hierarchy between coefficients of
absorption one can define the pre-critical region
by two conditions
\begin{itemize}
 \item
 by inequality
$$
F < F_{cm} -1
$$
where $F_{cm}$ is the minimal activation barrier among
different channels.
\item
by  requirement that
this region has to be continuous and the origin belongs to this
region.
\end{itemize}

 One can prove that in this region the quasi stationary
equilibrium
state takes place.
Here the absence of the hierarchy of  kinetic coefficients
plays the principal role.

Now one can investigate the form of the pre-critical region
in $\nu_i$ variables. It
looks like a star and the needles are going along the bottoms of
channels. Certainly due to restrictions $\nu_i \geq 0$ there is
only one quarter of a star.
In $\kappa, \{ \xi \} $ variables it looks like a brush.

If in every channel we put the value $F_c$
corresponding to this channel, the  shortest needle is the
main one. The shortest needle (in $\kappa, \xi$
plane) corresponds to the lowest barrier and, hence, it is the main
needle through
which the nucleation takes
place.

If the level $F_c$ is chosen as $F_{cm}$ and it is one and
the same for all channels then the main needle is the
longest one

To see the relaxation to the equilibrium distribution we need to
determine the minimal diameter of this star. It is given by the
following relation
$$
 - \kappa_{min} b_{g\ min} +  \kappa_{min}^{2/3}
= F_{cm} -1
$$
Here $b_{g\ min}$ is the minimal
value of $b_g$. So, if $| b_{g\ min} | $ does not go to
infinity, one can easily see the finite value of
$\kappa_{min}$ and the connection of channels.

The last consideration solves the problem of connection of channels
of nucleation. The problem was that the behaviour of channels near
the origin where the surface excesses can play the leading role was
unclear. So, one could not say whether the channels are connected
or no. Now the concrete position of channels near origin is not
important.

The only condition is the restriction on $b(\xi)$ - this function
can not go to $- \infty$ at some concentrations.

\subsection{Characteristic sizes of near-critical region.}

Consider the variables parallel to $\xi_i, \kappa$ and having the
scale of $\nu_i$. These variables are
$$
\nu_{par} \simeq \frac{\kappa  n }{   \gamma^{3/2} \sum_i^n v_i}
$$
$$
\nu_{i\ perp} \simeq \nu_{par} \xi_i
$$
Here it is supposed that all $v_i$ have the same order of values.
The total number of components $n$ is not supposed to be a big
parameter.

Then the halfwidths along $\nu_{par}$ and $\nu_{i\ perp}$ satisfy
the following estimates
$$
\Delta \nu_{par} \sim \kappa^{1/6} \Delta \nu_{i\ perp}
$$
$$
\Delta \nu_{par} \sim \kappa^{2/3}\sim \nu_{tot}^{2/3}
$$
$$
\Delta \nu_{perp} \sim \kappa^{1/2}\sim \nu_{tot}^{1/2}
$$

The time of establishing of the stationary state along $\nu_{par},
\nu_{i\ perp}$ is given by
$$
t_{r \ par} sim (
\frac{W^+}{\Delta^2  \nu_{par}})^{-1}
$$
$$
t_{r \ i perp} \sim (
\frac{W^+}{\Delta^2  \nu_{i\ perp}} )^{-1}
$$
Here all kinetic coefficients of absorption are supposed to have
one and the same order of value which is marked by $W^+$.

Then we come to the following strong inequality
$$
\frac{t_{r \ par} }{t_{r \ i perp}} \sim
\kappa^{1/3}
\gg 1
$$
This equation states the hierarchy in the near critical region.
Earlier this hierarchy was established in \cite{PhysicaA05} for the
situation without surface excesses. Here it is done for the
presence of the surface excesses.

To see this property
the main effort was spent to show the similarity of
forms of the free energy relief. Then one can come to the hierarchy
rather automatically.

The mean characteristic time $t_u$ to overcome the near-critical
region for the embryo at the bottom the channel at the boundary of
the near-critical and pre-critical regions has the order of $t_{r\
par}$
$$
t_u \sim t_{r\ par}
$$

Then we come to the following strong inequality
$$
\frac{t_{u} }{t_{r \ i perp}} \sim
\kappa^{1/3}
\gg 1
$$
It means that along $\nu_{i\ perp}$ there is a quasi equilibrium.

\subsection{Advantages of hierarchy}

On the base of hierarchical inequalities one can see that along
$\nu_{i\ perp}$ or $\xi_i$ there is quasi equilibrium. Then the
distribution function $n(\{ \nu_i \} ) $ which can be transformed
into $n( \nu_{par}, \{ \nu_{i\ par} \})$ can be presented as
$$
n( \nu_{par}, \{ \nu_{i\ par} \}) = N_{par}(\nu_{par}) n_{eq}(\{
\nu_{i\ perp} \})
$$
where $N_{par}$ plays the role of the amplitude of the known
equilibrium distribution and $n_{eq}$ is given by
$$
n_{eq} (\{ \nu_{i\ perp} \}) \sim
\exp( - F(\nu_{par} , \{ \nu_{i\ perp} \}) )
$$
or more convenient
$$
n_{eq} (\{ \nu_{i\ perp} \}) \sim
\exp(F(\nu_{par} , \{ \nu_{i\ perp} \})
-  F(\nu_{par} , \{ \nu_{i\ perp\ b} \}))
$$
where ${b}$ marks  the coordinate of the bottom of the
channel.

Then there remains only the task to determine the amplitude
$N_{par}$. This is a simple one-dimensional problem of
nucleation. One can easily solve it.

Reduction the problem of nucleation to the one dimensional case
allows to solve more complex situations. At first one can see that
when the characteristic width of equilibrium distribution seriously
changes it leads to the change of the effective free energy in
$N_{par}$. Really the effective free energy looks like
$$
F_{eff} = F - \ln \Delta_{eq} \nu
$$
where
$$
\Delta_{eq} \nu = (\sum_{\nu_{perp}} n_{eq} (\nu_{perp}))^{-1}
$$

In the majority of cases the summation in the last formula can be
replaced by integration
$$
\Delta_{eq} \nu =
(\int_{-\infty}^{\infty} n_{eq} (\nu_{perp}) d\nu_{perp})^{-1}
$$
Here the region of integration is formally put to an infinite one,
actually one has to integrate over the region
near the bottom of the channel where $n_{eq}$ is
essential.

The further simplification is the following: one can take the last
integral in square approximation for
the equilibrium distribution:
$$
n_{eq} (\{ \nu_{i\ perp} \}) \sim
\exp( - F_b)
\prod_j
\exp( - \frac{\partial^2 F(\nu_{par} , \{ \nu_{i\ perp} \})}
{2\partial \nu_{j\ perp}^2 }|_{\nu_{i\ perp} =\nu_{i\ perp\ b}}
(\nu_{j\ perp} -\nu_{j\ perp\ b})^2)
$$
This allows to take integrals explicitly.

Then the effective free energy is given by
$$
F_{eff} = F - \sum_j \ln \frac{\pi^{1/2}}{\sqrt{
\frac{\partial^2 F(\nu_{par} , \{ \nu_{i\ perp} \})}
{2\partial \nu_{j\ perp}^2 }|_{\nu_{i\ perp} =\nu_{i\ perp\ b}}}}
$$
Later one has to solve one-dimensional nucleation problem with the
effective free energy instead of the initial free energy.
As it will be shown later by demonstration of the plausible
derivation of Reiss' formula one has to be very attentive
at this step.

Ordinary in the near-critical region
the value $\Delta_{eq} \nu$ is constant
and there is no peculiarities in behavior of $F_{eff}$. Certainly,
in the square approximation the is an explicit solution of
Stauffer. But the approach
based on hierarchy leads to final analytical results in
more complex and may be exclusive cases. Really, here the square
approximation was taken only as an illustration.

One has to clarify the place of the presented approach in the task
to determine the nucleation flow. Ordinary  to
justify the total square approximation in the near critical zone
and to use the Langer-Stauffer's approach one has to adopt some
approximations including
the smooth behavior of the derivative of $b_g$ near the
bottom of the channel. But there is no clear evidence of the
regular behavior of $b_g$ near the bottom.
 So, the approach based on the hierarchy is
preferable.

On the base of hierarchy one can
also see many interesting and important
facts:
\begin{itemize}
\item
At first we see that
the quasi-unary condensation can not be described
in terms of the square approach. A direct transformation of the
formulas appeared in the Langer-Stauffer's approach does not lead
to the formulas of the unary nucleation. This occurs because
inevitably the square approximation has to be violated. So, we come
to {\bf
the impossibility of description of the quasi-unary nucleation
in terms of the standard  Stauffer's binary nucleation approach.}
\item
The next consequence of general results is the impossibility of
situation
of the inverse direction
proposed by Zisterman-Berezhkovskii, where \cite{Berej} the
Stauffer's approach meets difficulties. Really, now it is clear
that valleys have to be directed to the origin, but not at the
perpendicular direction as it is supposed in
the consideration of
Zisterman and Berezhkovskii.
\end{itemize}

One has to mention that the thermodynamics is rather formal and
can
give essential corrections
to the initial variant of the theory
if some other expressions for chemical
potentials and surface energy are taken. Certainly, these
expressions have to be the matter of discussion. But one can not
deny the possibility to come to the situation where the square
approximation is not suitable and one has to follow the approach
suggested here.

Here we suppose that these expressions are already given. They are
some external information for the theory developed here.

\section{Stauffer's and Reiss' solutions}

The main goal in the investigation of
the multicomponent nucleation is to
get essential corrections in comparison with the already known
approaches. For this purpose we shall examine the formulas of
Stauffer and Reiss for the nucleation rate.

\subsection{
Kinetic equation }

Consider the
binary case.
Introduce the Reiss' variables $x$, $y$ as the variables when the
free energy in the critical region has the form
$$
F = F_c -x^2 +y^2 ,
$$
where $F_c$ is the free energy in the saddle point. These variables
can be obtained from $\nu_1$, $\nu_2$ by rotation and
rescaling\footnote{ May be some part of the Lorenz transformation
with an arbitrary parameter has been made. So, these variables
aren't completely fixed.}.

Instead of rotation and rescaling it is more convenient to
introduce the separated variables directly. The variables $\kappa$,
$\xi$ are the stable and unstable ones. One can come to
$$
\frac{\partial^2 F}{\partial \xi \partial \kappa} =
-  \frac{d b_g(\xi) }{ d \xi}           ,
$$
which is vanished in the saddle point. It means that the square
form of the free energy in $\kappa, \xi$ variables looks like
$$
F = - A (\kappa - \kappa_c)^2 + B (\xi - \xi_c)^2 + F_c
$$
without the cross term. Here $A$ and $B$ are some positive
constants
$$
A = -(\frac{\partial^2 F(\kappa, \xi)}{2\partial \kappa^2} )_c
\
\
\
B = (\frac{\partial^2 F(\kappa, \xi)}{2\partial \xi^2} )_c
$$
Then in the coordinates
$$
\tilde{x} = {\sqrt{A}}({\kappa - \kappa_c})
\
\
\
\tilde{y} = {\sqrt{B}}({\xi - \xi_c})
$$
one gets
$$
F = F_c -\tilde{x}^2 + \tilde{y}^2
$$
Now we shall seek for the similar variables obtained by the linear
transformations.

The variables $x$, $y$ can be obtained from $\nu_1$, $\nu_2$ by the
linear transformation
$$
x = c_{11} (\nu_1 -\nu_{1c}) + c_{12} (\nu_2 -\nu_{2c}) ,
$$
 $$
y = c_{21} (\nu_1 -\nu_{1c}) + c_{22} (\nu_2 -\nu_{2c})
$$
(which isn't orthogonal) with the known coefficients
$$
c_{11} = [-\frac{1}{2} (\frac{\partial^2 F}{\partial
\kappa^2})_c]^{1/2} (\frac{\partial \kappa}{\partial \nu_1})_c ,
$$
$$
c_{12} = [-\frac{1}{2} (\frac{\partial^2 F}{\partial
\kappa^2})_c]^{1/2} (\frac{\partial \kappa}{\partial \nu_2})_c ,
$$
$$
c_{21} = [\frac{1}{2} (\frac{\partial^2 F}{\partial
\xi^2})_c]^{1/2} (\frac{\partial \xi}{\partial \nu_1})_c ,
$$
$$
c_{22} = [\frac{1}{2} (\frac{\partial^2 F}{\partial
\xi^2})_c]^{1/2} (\frac{\partial \xi}{\partial \nu_2})_c
 .
$$
The variables $x,y$ practically coincide with $\tilde{x}$,
$\tilde{y}$. The difference  has an order of a small parameter.

The estimates for coefficients $c_{11}, c_{21}, c_{12}, c_{22}$ are
$$
c_{11}
\sim
\kappa^{-2/3}_c               ,
$$
$$
c_{12}
\sim
\kappa^{-2/3}_c                ,
$$
$$
c_{21}
\sim
\kappa^{-1/2}_c                 ,
$$
$$
c_{22}
\sim
\kappa^{-1/2}_c                  .
$$

The estimates
$$
\Delta \kappa \sim \kappa^{2/3} \sim \nu_{tot}^{2/3} \sim
\Delta \nu_{par}
$$
$$
\Delta \nu_{perp\ i}  \sim \kappa^{1/2} \sim \nu_{tot}^{1/2}
$$
in positive powers of a big parameter $\kappa$ (or $\nu_{tot}$)
allows to use the Fokker-Planck's approximation.

In the Fokker-Planck's approximation the kinetic equation for the
distribution function $n$ can be written in the following form
$$
\partial_t n(\nu_1, \nu_2)
=
W_1 \partial_1 [ n \partial_1 F + \partial_1 n ] + W_2 \partial_2 [
n \partial_2 F + \partial_2 n ] ,
$$
where $W_1$, $W_2$ are the kinetic coefficients, i.e. the numbers
of the first sort molecules and the second sort molecules which are
absorbed by the embryo in the unit of time. Here
$$\partial_1 \equiv \partial / \partial \nu_1 ,
\ \ \ \ \ \partial_2 \equiv \partial / \partial \nu_2
$$
and $\partial_t \equiv \partial / \partial t
$.
The differentiation on the number of the molecules of the given
sort in marked by the index near the symbol of the partial
differentiation.

Now we rewrite the kinetic equation in the variables $x$, $y$. Note
that
$$
\partial_1 = c_{11} \partial_x + c_{21} \partial_y
$$
$$
\partial_2 = c_{12} \partial_x + c_{22} \partial_y
$$
where $\partial_x = \partial / \partial x$ and $\partial_y =
\partial / \partial y$.

The distribution $n(x,y)$ is proportional to the distribution
$n(\nu_1 , \nu_2)$ with coefficient $\partial(\nu_1,
\nu_2)/\partial (x,y)$ and one has to take   this difference
into account in final calculations.
In the near-critical region the coefficients
of kinetic equation are approximately
constants.

To simplify the treatment one can use notations
$$
\partial_{x1} = c_{11} \partial _x , \ \ \
\partial_{x2} = c_{12} \partial _x , \ \ \
\partial_{y1} = c_{21} \partial _y , \ \ \
\partial_{y2} = c_{22} \partial _y . \ \ \
$$
Then one can get the equation
\begin{eqnarray}
\partial_t n  =
W_1 (\partial_{x1}+\partial_{y1}) [ n(\partial_{x1}+\partial_{y1})F
+ (\partial_{x1}+\partial_{y1}) n ]
\nonumber
\\
\nonumber
+ W_2 (\partial_{x2}+\partial_{y2}) [
n(\partial_{x2}+\partial_{y2})F + (\partial_{x2}+\partial_{y2}) n ]
\end{eqnarray}
Since the structure of terms like $
n(\partial_{x1}+\partial_{y1})F$ coincide with the structure of
$(\partial_{x1}+\partial_{y1}) n$ one can simply miss the last term
and reconstruct it in the final expressions. Then
$$
\partial_t n =
K_1 \partial_x (n\partial_x F + \partial_x n) + K_2 [ \partial_x (n
\partial_y F + \partial_y n) + \partial_y (n
\partial_x F + \partial_x n ) ] + K_3 \partial_y (n \partial_y F +
\partial_y n )
$$
where
$$
K_1 = W_1 c_{11}^2 + W_2 c_{12}^2
$$
$$
K_2 = W_1 c_{11} c_{21} + W_2 c_{12}c_{22}
$$
$$
K_3 = W_1 c_{21}^2 + W_2 c_{22}^2
$$

To stress the hierarchy one can introduce the coefficients
$$
R = K_1 , \ \ k = -\frac{K_1}{K_2}, \ \ q = \frac{K_3 K_1}{K_2^2}
$$

Then finally
$$
\partial_t n(\nu_1, \nu_2)
=
R [ \partial_x [ n \partial_x F + \partial_x n ]
-
k^{-1} [\partial_x [ n \partial_y F + \partial_y n ] +
\partial_y [ n \partial_x F + \partial_x n ]]
$$
$$
+ k^{-2} q
 \partial_y [ n \partial_y F + \partial_y n ]
]
$$

For $R$, $k$, $q$ one can get the following expressions
$$
R = W_1 c_{11}^2 + W_2 c_{12}^2 ,
$$
$$
k =
-
\frac{
W_1 c_{11}^2 + W_2 c_{12}^2 }{ W_1 c_{11} c_{21} + W_2 c_{12}
c_{22} } ,
$$
$$
q = \frac{(W_1 c_{21}^2 +W_2 c_{22}^2 )(W_1 c_{11}^2 +W_2 c_{12}^2
)}{(W_1 c_{11} c_{21} + W_2 c_{12} c_{22})^2}
$$
The last coefficient can be also written as
$$
q = 1 + W_1 W_2 ( \frac{c_{11} c_{22} - c_{12} c_{21}} {W_1 c_{11}
c_{21} + W_2 c_{12} c_{22}} )^2 .
$$

The value of $R$ isn't important because it can be changed by the
time rescaling. One can see the estimate
$$
k \sim \nu_c^{-1/6}
$$
which shows that $k$ is a small parameter. The scale of $q$ is
arbitrary, but one can outline situations where $q-1
\ll 1 $.

The boundary conditions for the last equations are the following
\begin{eqnarray}\label{bound}
 n/n^e = 1 \ \ \ \ \  x \ll -1 \  \ \ \ \ -\infty < y  < \infty  ,
 \nonumber
 \\
 \\
 n/n^e = 0 \ \ \ \ \  x \gg 1 \  \ \ \ \ -\infty < y  < \infty
 \nonumber
\end{eqnarray}

The plausible but not rigorous consideration
corresponding to the solution proposed by Reiss is the
following one
\begin{itemize}
\item
The main operator of kinetic equation is the last term in r.h.s.

\item
It
ensures the relaxation over the stable variable and the kinetic
equation becomes the one dimensional one.

\item
The consideration of the
evolution only over the unstable variable leads to the reduction of
the kinetic equation to
$$
\partial_t n(\nu_1, \nu_2)
=
R
 \partial_x [ n \partial_x F + \partial_x n ]
$$

\end{itemize}

The solution of the last equation leads to the results of Reiss.
But in the cited paper of Reiss the hierarchy was not observed.
Hence, the analysis there was less plausible.

\subsection{
The influence on the characteristics of the process}

One needs the transformation of kinetic equation which conserves
the boundary conditions, since the variables in the boundary
conditions (\ref{bound}) are already separated. This transformation
is the Lorenz' transformation.

Introduce the Lorenz' transformation via formulas
$$
\psi = \frac{x+\alpha y}{\sqrt{1-\alpha^2}}
 , \ \ \ \ \
\eta = \frac{y+\alpha x}{\sqrt{1-\alpha^2}}
$$
This transformation
 conserves the form of the free energy in the critical
region:
$$
F = F_c - \psi^2 + \eta^2
$$

The kinetic equation is transformed to
$$
\partial_t n(\nu_1, \nu_2)
=
R (1-\alpha^2)^{-1} k^{-2} [ [(k-\alpha)^2 + \alpha^2 (q-1)]
\partial_{\psi} [ n \partial_{\psi} F + \partial_{\psi} n ]
-
$$
$$
[(k-\alpha)(1-k\alpha)-\alpha(q-1)] [\partial_{\psi} [ n
\partial_{\eta} F + \partial_{\eta} n ] +
$$
$$
\partial_{\eta} [ n \partial_{\psi} F + \partial_{\psi} n ]]
+ [(1-k\alpha)^2+q-1]
 \partial_{\eta} [ n \partial_{\eta} F + \partial_{\eta} n ]
] .
$$

Parameter $\alpha$ which has the absolute value less than $1$ has
to be chosen to vanish the cross term. The equation for the choice
of $\alpha$ is the following
$$
(k-\alpha)(1-k\alpha) = \alpha (q-1)
$$
Then
$$
\partial_t n(\nu_1, \nu_2)
=
A
\partial_{\psi} [ n \partial_{\psi} F + \partial_{\psi} n ]
+ C
 \partial_{\eta} [ n \partial_{\eta} F + \partial_{\eta} n ]
$$
where
$$
A = \frac{R}{k^2} (1-\alpha^2)^{-1} [(k-\alpha)^2 + \alpha
(k-\alpha) (1 - k \alpha)]
$$
$$
C =
\frac{R}{k^2} (1-\alpha^2)^{-1}
[(1-k\alpha)^2 + \frac{(k-\alpha)(1-k\alpha)}{\alpha}]
$$

The parameter of the Lorenz' transformation is given by
$$
\alpha = \frac{1}{2k} [ k^2 + q - \sqrt{(k^2 + q)^2 - 4 k^2 }]  .
$$
After the decomposition at small $k$ one can come to
\begin{equation}\label{al}
\alpha = \frac{1}{q} k                                           .
\end{equation}
in the leading term.
One can see that it is small. So, it is difficult to see the effect
of the Stauffer's consideration on the direction of the flow. But
one can not directly put $\alpha = 0$ because there is a small
parameter $k$.
Expression for $A$ will be
\begin{equation}\label{A}
A = R \frac{q-1}{q} .
\end{equation}
The ratio $1/q$ is not small.
So the correction to the Reiss' formula is
essential. The direct substitution $\alpha = 0 $ leads to
$$
A|_{\alpha = 0} = R
$$
which is the Reiss' result and it is not precise.

\subsection{Some consequences for the binary nucleation}

The question to discuss here is the rate of the deviation of the
Reiss' formula for the nucleation rate from the analogous result of
Stauffer.

In the derivation of the expression for $q$ no suppositions about
$W_1$ and $W_2$ have been made. At first the situation with the
moderate ratio $W_1/W_2$ will be discussed.

As far as
\begin{equation} \label{ll}
\frac{\partial \xi }{\partial \nu_1}
= \frac{\partial (1-\xi)}{\partial
\nu_2} = - \frac{\partial \xi}{\partial \nu_2}
\end{equation}
we see that the
partial cancellation can take place in expression for $q$
only in $$ {W_1 c_{11} c_{21} + W_2 c_{12} c_{22}}$$ but not in
$$c_{11} c_{22} - c_{12} c_{21}$$ So $q$ is big enough to lead to
result near the Reiss' formula $A=R$. This shows that the Reiss'
formula is not so bad although it is not a true result.

The precise coincidence of Reiss' and Stauffer's results takes
place when $q=\infty$, i.e. when
$$
W_1 c_{11} c_{21} + W_2 c_{12} c_{22} = 0
$$
The last relation taking into account (\ref{ll}) can be rewritten
as
$$
W_1 \frac{\partial \kappa}{\partial \nu_1} = W_2 \frac{\partial
\kappa}{\partial \nu_2}
$$
In the rough approximation corresponding to:
\begin{itemize}
\item
 the capillary
approximation itself,
\item
the Gibbs-Duhem' equation in the capillary
approximation
\item
 the negligible dependence of $v_i$ on $\kappa$ in capillary
approximation
\end{itemize}
one can see
that the last relation transforms to
$$
W_1 v_1 = W_2 v_2
$$
where $v_i$ is the volume per molecule in a liquid phase. This
condition is the condition of precise applicability of the Reiss'
result. It differs from condition
$$
W_1 = W_2
$$
announced in paper \cite{Berej} analyzing the theory of Stauffer.

It is clear that the last condition is wrong which opens a question
of the formal validity of the Stauffer's derivation.
Really, formally regarding one
molecule of the first substance as several particles, one can
attain the applicability of Reiss' result by such an artificial
way. For condition $W_1v_1 = W_2 v_2$, this trick fails.

Let us extract the conditions when $A$ essentially differs from
$R$. It can be only when
$$q \approx 1$$

The last condition can be satisfied only when $W_1 \ll W_2$,
$W_1
\gg W_2$.
Namely, this situation occurs when there is the rapid component.
 {\it {
The essential variation of the nucleation rate in comparison with
the Reiss' formula is possible only under the hierarchy of the
kinetic coefficients. }}  This situation requires a separate
analysis.

As an illustration
here we shall show the result in a square approximation
of the free energy, although one can analytically prove that the
existence of a rapid component throws the main nucleation flow away
from the near-critical region and another approximations for the
free energy have to be used.

Under the hierarchy one can see the evident rapid component and
formulas can be simplified. The simplification can be made also
directly in the final formulas and the expression for $A$
$$
A = W_2 \frac{(c_{11}c_{22} - c_{12} c_{21})^2} {c_{21}^2}
$$
is proportional to $W_2$. Then
$$
A = W_2 (- \frac{1}{2} \frac{\partial^2 F}{\partial \kappa^2}) [
\frac{\partial \kappa}{\partial \nu_1} +
 \frac{\partial \kappa}{\partial \nu_2}]^2
$$

In the further considerations of this section this simplification
is not used.

\subsection{
Conclusions based on hierarchy}

In the post critical region one can assume the derivative on the
unstable variable to be  locally a constant and reduce the kinetic
equation to
$$
  \partial_t n =
R [ \partial_x (\partial_x + h) - k^{-1} ( \partial_x ( \partial_y
+ 2 y) + \partial_y (\partial_x + h) ) + k^{-2} q \partial_y (
\partial_y + 2 y) ] n ,
$$
where $h$ is the constant coefficient corresponding to the first
derivative on the unstable variable and the values of $R$,$k$, $q$
are changed since  the derivatives are taken now in the local
current point. Renormalize the scale over the unstable variable as
to put $h=1$. Certainly, the hierarchy takes place after the
renormalization.

We are interested in the stationary solution and shall seek it in
the form
\begin{equation}\label{p}
n = Q(x) \exp(-(y-y_0)^2)
\end{equation}
with the constant mean value $y_0$ and some function $Q(x)$. The
derivative $d Q(x) / dx $  can be neglected. Then
$$
[-k \partial_y h + q \partial_y (\partial_y + 2y)] n = 0
$$
 For $y_0$
one can get taking into account
$$
\partial_y  \exp(-(y-y_0)^2) = - 2 (y-y_0) \exp(-(y-y_0)^2)  ,
$$
$$
\partial_y (\partial_y + 2 y ) \exp(-(y-y_0)^2) = - 4 y_0 (y-y_0)
\exp(-(y-y_0)^2)
$$
the following relation
\begin{equation}\label{yo}
\frac{k h}{q} =  2 y_0                                     .
\end{equation}
So the solution is obtained.

Consider this solution. We see that the deviation of the rapid
parameter is small also in the post critical region and the
possible hidden
parameter can not be extracted.

Due to the slope of the free energy surface on $\kappa$ the minimum
of the free energy in the cross section depends on the slope of
this cross section. But since the slope of the free energy
surface on $\kappa$ is small the deviation of the minimum is small
also. This deviation can be considered as the deviation of the mean
value of the rapid variable and leads to the absence of the
possibility to extract this variable in the post critical region
also.

The analogous method can be applied also for the near-critical
region. In the near-critical region one can make the substitution
$$
n = P(x) \exp(-(y-y_0(x))^2) ,
$$
where $y_0$ is now the function of $x$. One can determine $y_0$
according to
$$
(\partial_x - 2 x) n = - J_x ,
$$
there the r.h.s. is constant. Then
\begin{equation}\label{lll}
 \partial_x  n = - J_x + 2 x n  .
 \end{equation}
The linear character of the last equation ensures the linear
dependence of $y_0$ on the unstable variable. As far as the flow is
reciprocal to the halfwidht (along the trajectory $y_0$) one can
get the equation on the flow. The linear dependence
of $y_0$ on $x$
ensure the linear character of the transformation which is
analogous to the Lorenz' transformation.

This way of considerations can be applied to the more general
situations without the square form of the free energy. Then the
trajectory isn't the straight line and the solution is some
approximation based on the hierarchy.

The last question to solve is a real position of the near-critical
region.

When the deviation of the flow from the steepens descent situation
is essential there is the danger to violate the square form of the
free energy. The boundary conditions for kinetic equation in the
critical region  in reality have to be observed at
$$ n/n^e = 1 \ \ \ \ \   x \sim -1 \ \ \ \ -1 < y < 1  , $$
$$ n/n^e = 0 \ \ \ \ \  x \sim 1 \ \ \ \ -1< y < 1  . $$
After the Lorenz' transformation
$$ n/n^e = 1 \ \ \ \ \   \psi \sim -1 \ \ \ \ -1 < \eta < 1  , $$
$$ n/n^e = 0 \ \ \ \ \  \psi \sim 1 \ \ \ \ -1< \eta < 1  . $$

Rigorously speaking one has to put the equilibrium
conditions at the line where $F=F_c-1$ which is invariant
to Lorenz' transformation. But actually, to ensure the
finite relaxation time and the constant values of kinetic
coefficients one has to cut-off the tails and to go to the
boundary of the reduced near-critical region. But this
boundary is not invariant to Lorenz' transformation.

The last  definition
of the boundary conditions has to be considered as the main one.

But here
the reduced near-critical
region is stretched along one of the lines $F=F_c$
where the transition occurs.
The square approximation in such stretched region can be invalid.

\subsection{Conclusions}

The main new results of the consideration made above are the
following:
\begin{itemize}
\item
The hierarchy of terms in kinetic equation is shown. Earlier the
hierarchy was observed only for halfwidths of the near critical
region \cite{PhysicaA05}.
\item
The plausible way to derive the Reiss' formula was demonstrated.
Since this formula is wrong, this deviation
demonstrates the impossibility to
neglect in kinetic equation all terms except the main one.
\item
The moderate value of the error made by Reiss is established.
Earlier there was a strong conviction that the error of the Reiss'
approach can be enormous, which was illustrated by numerical
examples in \cite{Stauffer}. Now it is clear that the big error can
be only in the cases of strong hierarchy between kinetic
coefficients ($W_1 \gg W_2, W_2 \gg W_1$) when the nucleation flux
goes mainly far from the saddle point.
\item
The simplified relations for $\alpha$ (see (\ref{al})) and for the
nucleation rate (see equation (\ref{A}) for $A$) have been derived.
\item
The super-critical region is studied and the expression for the
distribution function over the stable variable (\ref{p}),
(\ref{yo}) in this region is derived.
\end{itemize}

One can see that the precise result is rather complex. It can not
be achieved by  a simple superposition of naive solutions based on
hierarchy. One has also to mention that even in hierarchy $W_1 \gg
W_2$ the result differs from the naive one.

But the main result is the absence of the really important
corrections in comparison with a
naive approach.
All obtained corrections are rather small and mainly
less than the microscopic corrections in real situations. Below, we
shall seek  essential corrections in the case of hierarchy.

\section{Nucleation rate in the situation with hierarchy}

The case of hierarchy certainly requires a special consideration
going outside the local approximations in the neighborhood of a
saddle point of the embryos free energy.

There are many substances for which
the densities $n_{\infty i}$ have the
different orders of the values. For example,
$$
\frac{n_{\infty H_2O}}{n_{\infty H_2SO_4}} > 10^5
$$
in the everyday thermodynamic conditions.

Assume that there are two groups of substances: the substances
with a slow exchange and the substances with a rapid exchange.
Suppose
$$
W_a^+ \ll W_b^+$$
 The
components of the first group will be marked by the index "a" and
the components of the second group will be marked by the index "b".
At first we shall consider the situation of two components and
later the generalization will be evident.

Here the variable $\tilde{\upsilon}$ is the following one
\begin{equation}
\tilde{\upsilon} = \sum_a v_{la}\nu_a
\ \ .
\end{equation}

\subsection{Direction of a flow}

Extract the conditions when the flow is parallel to $\nu_b$.
 We construct a simple model
which will show some estimates.

The quantity of the embryos at the bottom with a fixed
slow component can be estimated
from above by
$$
N_{above} = \Delta \nu n_0\exp(-F_b)
$$
where
$F_b$ is the free energy at the bottom
$\Delta \nu$
is the effective width of the bottom.
 The normalizing
factor $n_0$ in some situations of the overcoming of the few
activation barriers can differ from the standard one. That's why we
keep a special definition for this factor.

The quantity of the embryos in the critical region which change the
number $\nu_a$ in the unit of time is
$$I_A = W_a^+\Delta\nu n_0\exp(-F_b)$$

The flow over the ridge is $J_s$.
So, the necessary condition is the
following
\begin{equation}\label{3.2.6.2}
W_a^+ \Delta\nu n_0\exp(-F_b) \ll J_s
\ \ .
\end{equation}

One can adopt for $J_s$ the following expression
\begin{equation}
J_{sb} = W_b^+ n_0 \exp(-F_t)/\Delta\nu \pi^{1/2}
\ \
\end{equation}
where $F_t$ is the free energy at the top of the ridge
and put $\Delta \nu$ as
\begin{equation}
\Delta \nu =
(\frac{\partial^2F(\nu_a,\nu_b)}{2
\partial\nu^2_b})^{-1/2}|_{\nu_a=\nu_{ac},\nu_b=
\nu_{bc}} \ \ .
\end{equation}
It is necessary that the transition occurs earlier than the
near-critical region is attained. Then it is possible to
put
$$
F_b = F_c -1
$$
at the boundary of the near-critical region. At the same
boundary one can also put
$$
F_t = F_c +1
$$
The inequality (\ref{3.2.6.2}) comes to
\begin{equation}\label{3.2.6.5}
W_a^+ \ll \frac{W_b^+}{\exp(2)(\Delta\nu)^2 \pi^{1/2}} \ \ .
\end{equation}

Practically the same condition can be obtained by the comparison of
the characteristic time between the transitions of the embryo along
$\nu_a$ which is $$t_{tr} \sim
(W_a^+)^{-1}$$ and the time of the relaxation in
the bottom
$$
t^s = \frac{(\Delta\nu_b)^2}{W_b^+} \ \ .
$$

\subsection{The normalizing factor}

Here we shall see that there is no equilibrium distribution
in the whole pre-critical region.

Extract the condition when there will be the equilibrium
distribution at the level with the fixed $\nu_a$
of the pre-critical
region.
 The quasi
equilibrium distribution has the form
\begin{equation}
n = n^q = n^q_0 \exp(-F(\nu_a,\nu_b))|_{\nu_a = const}
\ \ .
\end{equation}

The normalizing factor $n^q_0$ differs from the standard
normalizing factor because there is an equilibrium along the band
but there is no equilibrium between bands.

To establish the equilibrium it
is sufficient to have the intensity of the contact between the
neighbor bands greater than the intensity of the overcoming over
the activation barrier.
So, it is necessary to determine the height of the
activation barrier. Choose as $\nu_b$ the value of $\nu_{be}$,
corresponding to the minimum of the free energy at the band
\begin{equation}
\nu_{be} \ : \ min_{\nu_b} F(\nu_a,\nu_b) =
F(\nu_a,\nu_b) \ \ .
\end{equation}
Then the intensity of the contact can be estimated by
$W^+_an_0\exp(-F(\nu_a, \nu_{be})$.

 One can due to
    (\ref{3.2.6.5})
assume that the transition to the post critical region occurs along
 $\nu_b$,
i.e. inside the band\footnote{ The value $\Delta
\nu_b$
depends on $\nu_a$ weakly.}. Beside $\nu_{be}$ one can
introduce $\nu_{bx}$ as the point inside the band where the free
energy has the maximum
\begin{equation}
max_{\nu_b} F(\nu_a,\nu_b) = F(\nu_a,\nu_{bx})
\equiv F_x(\nu_a) \ \ .
\end{equation}

Under the square approximation the transition along $\nu_b$ can not
occur because this variable is the stable one. Then $\nu_{b \ x}$
can not be defined. But if the component $\nu_b$ is supersaturated
over the pure plane liquid then the condensation into the pure
liquid is possible and $\nu_{b \ x}$ must exist.
This shows that the square approximation can not be used here.

The transition into the super critical region can occur under the
arbitrary $\nu_a$. But the probability of such transition is very
low for all $\nu_a$ when $\exp(-F_x(\nu_a))$
strongly differs from
$\exp(-F_c)$, i.e. out of the critical region. But it can be
greater than the intensity to come to the next band. The intensity
of the establishing of the equilibrium (not the quasi equilibrium)
at the next band\footnote{ The intensity of transition to the next
band.} is less than the intensity of the transition over the ridge.
 This intensity is given by
\begin{equation}
J = J_s = n^q_0 \exp(-F_x(\nu_a))W^+_{bx} / \Delta_x \nu_b \pi^2
\ \ ,
\end{equation}
where
\begin{equation}
W^+_{bx} = W_b^+(\nu_a,\nu_{bx});
\ \ \ \
\Delta_x\nu_b =
(\frac{\partial^2F(\nu_a,\nu_b)}{2\partial\nu_b^2})^{-1/2}|_{\nu_b
= \nu_{bx}}
\ \ .
\end{equation}

There is no need to establish the equilibrium along the whole band
with the small $\nu_a$. The value of $\nu_{bx}$ for small $\nu_a$
can be very big, the barriers of the nucleation
can be very high, but it
is necessary to have the equilibrium only near the bottom, i.e. at
$\nu_b$ near to $\nu_{be}$. The establishing of the equilibrium
along the whole pre-critical region of the band is necessary only
for the bands where the intensity of the transition to the post
critical region is essential (comparable with the intensity of the
transition between the bands). According to the previous
considerations there is the quasi equilibrium along such bands.

Introduce the number of embryos in the band
\begin{equation}
N(\nu_a) = n^q_0 \exp(-F(\nu_a,\nu_{b\ e})) \Delta_e \nu_b
\ \ ,
\end{equation}
where
\begin{equation}
\Delta_e \nu_b =
\sum_{\nu_b =0}^{\nu_{bx}}
\exp(-F(\nu_a,\nu_b)+F(\nu_a,\nu_{be}))
\ \
\end{equation}
has a sense of characteristic width.
The last formula in a continuous limit can be transformed to
\begin{equation}
\Delta_e \nu_b =
\int_0^{\nu_{bx}}
\exp(-F(\nu_a,\nu_b)+F(\nu_a,\nu_{be}))d\nu_b
\ \ .
\end{equation}

At the ends of the interval of integration the equilibrium
distribution can be violated but there the subintegral function
goes to zero. As far as $\exp(-F)$ as function of
 $\nu_b$
is rather sharp near the maximum then the number of the embryos
going from the band with $\nu_a$ to the band with $\nu_a-1$ can be
approximated by $W^-_a(\nu_a,\nu_{be})N(\nu_a)$. The number of the
forward transitions is $W_a^+(\nu_a-1,\nu_{be})N(\nu_a-1)$. Then one
can write the balance equation at the band
\begin{eqnarray}
\frac{\partial  N}{\partial  t} = W_a^+(\nu_a-1,\nu_{be}) N(\nu_a-1)
+W_a^-(\nu_a+1,\nu_{be}) N(\nu_a+1) -
\nonumber \\
\\ \nonumber
W_a^+(\nu_a,\nu_{be}) N(\nu_a) - W_a^-(\nu_a,\nu_{be}) N(\nu_a) -
J(\nu_a) \ \ .
\end{eqnarray}
For $J(\nu_a)$ one can get
\begin{equation}\label{3.2.6.17}
J(\nu_a) = N \frac{W^+_{bx}\exp(F(\nu_a,\nu_{be}) - F_x(\nu_a)) }
{\Delta_e\nu_b \Delta_x \nu_b}
\ \ .
\end{equation}

One can see that the absence of the equilibrium distribution in the
whole
pre-critical region is the
characteristic feature  of the transition far from
the saddle point.

\subsection{Valley zone and ridge zone}

For every $\nu_a$ in the pre-critical region
 there will be $\nu_{be}$. The curve $\nu_{be}(\nu_a)$ will be
 called the valley in $\nu_a, \nu_b$ plane.

For every $\nu_a$ in the  region under consideration
 there will be $\nu_{bx}$. The curve $\nu_{bx}(\nu_a)$ will be
 called the ridge in $\nu_a, \nu_b$ plane.

Since there is a slope of the ridge
and the valley in $\nu_a$ direction it is necessary to specify the
set of variables.

In the set of variables $\kappa, \xi$ the channel of nucleation is
the straight analog of a valley. But the channel of nucleation
does
not coincide with the the valley in $\nu_a, \nu_b$ plane.

The line analogous to the ridge, i.e. the ridge in $\kappa, \xi$
plane will be the separation line defined as
$$
\frac{\partial F(\kappa, \xi) }{\partial \xi} = 0
$$
$$
\frac{\partial F(\kappa, \xi) }{\partial \xi} < 0
$$

The values at the
channel of nucleation here will be marked by the subscript $h$
and at the separation line the values will be marked by the
subscript $s$.

We see that effectively the flow is directed along $\nu_b$. The
problem to get $J(\nu_a)$ is purely a one dimensional
problem. So, in the
band $\nu_a = const$ there exists the valley $\nu_b \approx
\nu_{b\ e}$ zone and the ridge $\nu_b \approx \nu_{b \ x}$ zone.
Precise definitions are the following
\begin{itemize}
\item
The ridge zone in $\nu_b$ scale is determined by conditions
$$
F(\nu_a, \nu_b) \geq F(\nu_a, \nu_{bx})-1
$$
Certainly, $F(\nu_a, \nu_b) \leq F(\nu_a, \nu_{bx})$.
This zone has to be near the given ridge.
\item
The valley zone in $\nu_b$ scale is determined by conditions
$$
F(\nu_a, \nu_b) \leq F(\nu_a, \nu_{be})+1
$$
Certainly, $F(\nu_a, \nu_b) \geq F(\nu_a, \nu_{be})$. This
zone has to be near the given valley.
\end{itemize}

To find the value of the flow $J(\nu_a)$ one has to solve kinetic
equation in the ridge zone. To find the normalizing factor like it
was done in heterogeneous nucleation it is necessary to consider
the valley zone and to solve kinetic equation in this region.

The problem under consideration is the influence of the surface
excesses on the forms of the free energy in the ridge zone and the
valley zone. Fortunately some simplifying properties will be
established below which help to escape from the explicit inclusion
of surface excesses in the kinetic equation.

For the ridge zone these properties are the following
\begin{itemize}
\item
Define by the subscript $0$ the values without surface excesses
\item
In the ridge zone for arbitrary $s$ corresponding to the ridge zone
$$
F(\nu_a, \nu_{bx}+s) - F(\nu_a, \nu_{bx})
\approx
F(\nu_a, \nu_{bx0}+s) - F(\nu_a, \nu_{bx0})
$$
\end{itemize}

For the valley zone these properties are the following
\begin{itemize}
\item
In the valley zone for arbitrary $s$ corresponding to the valley
zone
$$
F(\nu_a, \nu_{be}+s) - F(\nu_a, \nu_{be})
\approx
F(\nu_a, \nu_{be0}+s) - F(\nu_a, \nu_{be0})
$$
\end{itemize}

One can analogously define the channel zone and the separation
zone.

\begin{itemize}
\item
The separation zone  is determined by conditions
$$
F(\kappa, \xi) \geq F(\kappa, \xi_{s})-1
$$
The value of $\kappa$ is fixed here.
Certainly, $F(\kappa, \xi) \leq F(\kappa, \xi_{s})$.
The separation zone has to be near the given separation
line.
\item
The channel zone  is determined by conditions
$$
F(\kappa, \xi) \leq F(\kappa, \xi_{h})+1
$$
The value of $\kappa$ is fixed here.
Certainly, $F(\kappa, \xi) \geq F(\kappa, \xi_{h})$.
The channel zone has to be near the given channel line.
\end{itemize}

One can analytically prove the following properties for
 the separation zone
\begin{itemize}
\item
In the separation zone
for arbitrary $s$ corresponding to the separation zone
$$
F(\kappa, \xi_{s}+s) - F(\kappa, \xi_{s})
\approx
F(\kappa, \xi_{s0}+s) - F(\kappa, \xi_{s0})
$$
\end{itemize}

One can analytically prove the following properties for
 the channel zone
\begin{itemize}
\item
In the channel zone
for arbitrary $s$ corresponding to the channel zone
$$
F(\kappa, \xi_{h}+s) - F(\kappa, \xi_{h})
\approx
F(\kappa, \xi_{h0}+s) - F(\kappa, \xi_{h0})
$$
\end{itemize}

The method of a proof of all these properties is quite analogous
to the already presented for the near-critical region.
These properties allow to solve kinetic equations in these
regions by some shift renormalizations and solutions in the
absence of the the surface excesses.

\subsection{Discrete case}

Consider the stationary solution. The last equations form the
system of algebraic equations. Note that the sufficient equations
are those where $W_a^+ N$ has the order of $J$. The equations with
$W_a^+ N \gg J$ can be taken into account by the boundary condition
$n= n^q = n^e$ for $\nu_a$ which is less than some $\nu_{amin}$,
where $J$ begins to be comparable with $W_a^+N$.
More precisely this question will be discussed later.

Formally one has to put this condition at $\nu \ll
\nu_{a min}$. Then one has to solve equations and to see
where the condition $n \approx n^e$ will be violated. It is
very easy to do having calculated $J(\nu_a)$ on the base of
$n^e$ to get $$n \approx n^e - \int_{-\infty }^{\nu_a}
J(\nu_a') d \nu_a'$$ or having expelled the unphysical
region
$$n \approx n^e - \int_{1}^{\nu_a}
J(\nu_a') d \nu_a'$$
 This will give the necessary estimate.

In the region
where $W^+_a N \ll J$ the solution is rather simple
\begin{equation}
n \ll n^e
\ \ .
\end{equation}
This condition will be seen automatically at some $\nu_a$
and since $n/n^e$ is a decreasing function of $\nu_a$ it
will take place later.
So, one has to investigate only few equations of the type
\begin{eqnarray}\label{3.2.6.19}
W_a^+(\nu_a-1,\nu_{be}) N(\nu_a-1) +W_a^-(\nu_a+1,\nu_{be})
N(\nu_a+1)
\nonumber
\\
\\
\nonumber
-W_a^+(\nu_a,\nu_{be}) N(\nu_a)
-W_a^-(\nu_a,\nu_{be}) N(\nu_a)
 = J(\nu_a)
\ \ .
\end{eqnarray}
The total flow is defined as
\begin{equation}
J_{int} = \sum_{\nu_a = \nu_{amin}}^{\nu_{amax}} J(\nu_a)
\ \ ,
\end{equation}
where $\nu_{amax}$ marks the upper boundary of the equations
sufficient for the consideration.

In the limit when there is only one sufficient equation\footnote{
Having attained $\nu_a$ all embryos come automatically to the super
critical region. Then it is possible to write the expression for
the transition on $\nu_b$.}
\begin{equation}
J_{int} = W_a^+
\int_0^{\nu_{bx}} n_0 \exp(-F(\nu_{a},\nu_b)) d\nu_b \ \ =
 W_a^+ N_{tot}(\nu_a)
\end{equation}
where the total number of droplets at $\nu_a$ is
$$
N_{tot}(\nu_a) =
\int_0^{\nu_{bx}} n_0 \exp(-F(\nu_{a},\nu_b)) d\nu_b
$$

The discrete
situation is the most frequent one. But namely this situation has
not been considered earlier.

\subsection{Differential model}

Consider the opposite situation when among (\ref{3.2.6.19}) there
are so many equations that it is difficult to solve the algebraic
equations. Then it is reasonable to come to the differential form.
The condition of the validity of the differential form coincides
with the condition of the big number of the essential equations.
Then
\begin{eqnarray}
J(\nu_a) = -\frac{\partial }{\partial \nu_a}
\{(W^+_a(\nu_a,\nu_{be}) -
W_a^-(\nu_a,\nu_{be})) N(\nu_a) \} +
\nonumber
\\
\\
\nonumber
\frac{\partial^2}{2 \partial\nu_a^2}
\{(W^+_a(\nu_a,\nu_{be}) +
W_a^-(\nu_a,\nu_{be})) N(\nu_a) \}
\ \ .
\end{eqnarray}

With account of (\ref{3.2.6.17}) one can get
\begin{eqnarray}
N
\frac{\exp(F(\nu_a,\nu_{be})-F_x(\nu_a))}
{\Delta_e\nu_b \Delta_x\nu_b} W^+_{bx}
=
\nonumber
\\
\frac{\partial}{\partial\nu_a}
\{(W_a^+(\nu_a,\nu_{be})
(1-\exp(\frac{\partial F(\nu_a,
\nu_{be})}{\partial\nu_a})))N(\nu_a)\} +
\\
\frac{\partial^2}{2\partial\nu_a^2}
\{(W_a^+(\nu_a,\nu_{be})
(1+\exp(\frac{\partial F(\nu_a,
\nu_{be})}{\partial\nu_a})))N(\nu_a)\}
\ \ .        \nonumber
\end{eqnarray}

One can note that
\begin{itemize}
\item
The hierarchy of the halfwidths of the near-critical region shows
that the quasi-unary nucleation in the square approximation in the
neighborhood of the saddle point is impossible.
\end{itemize}
 So, the change of
approximation to a linear one is absolutely necessary.
This conclusion is very essential  for further
consideration.

One can use the following approximations
\begin{equation}
J = J_0 \exp(cy) \ \ ,
\end{equation}
\begin{equation}
y=\nu_a - \nu_{a0} \ \ ,
\end{equation}
\begin{equation}
c = \frac{\partial
F(\nu_a,\nu_{be})}{\partial\nu_a}|_{\nu_a=\nu_{a0}}
- \frac{\partial F(\nu_a,\nu_{bx})}{\partial\nu_a}|_{\nu_a=\nu_{a0}}
\ \ ,
\end{equation}
\begin{equation}
J_0 = J(\nu_a)|_{\nu_a=\nu_{a0}} \ \ .
\end{equation}
It means that the linear approximation for $F(\nu_a, \nu_{bx}) -
F(\nu_a, \nu_{be})$ is adopted. The supposition made in this paper
radically changes from the supposition of Trinkaus. This
difference will be discussed in a special part of this paper.

One has to note that
$$
\frac{\partial F(\nu_a, \nu_{be})}{\partial \nu_a}
$$
differs from
$$
\frac{\partial F(\nu_a, \nu_{b})}{\partial \nu_a}
$$
and
$$
\frac{\partial F(\nu_a, \nu_{bx})}{\partial \nu_a}
$$
differs from
$$
\frac{\partial F(\nu_a, \nu_{b})}{\partial \nu_a} \ \ .
$$
When we use $
\frac{\partial F(\nu_a, \nu_{be})}{\partial \nu_a}
$
we
imply the differentiation along the bottom of a valley.
When we use $
\frac{\partial F(\nu_a, \nu_{bx})}{\partial \nu_a}
$
we
imply the differentiation along the top of a ridge.

Then one can get
\begin{equation}
I \exp(cy) N =
- W_a^+ (1-\epsilon) \frac{dN}{dy}
+ W_a^+ (1+\epsilon) \frac{d^2N}{2dy^2} \ \ ,
\end{equation}
where
\begin{equation}
I = \frac{ W_{bx}^+ \exp(F(\nu_{a0}, \nu_{be}) - F_x(\nu_{a0}))}
{\Delta_e\nu_b \Delta_x\nu_b}
\ \ ,
\end{equation}
\begin{equation}
W_a^+ = W_a^+(\nu_a,\nu_{be})
\ \ ,
\end{equation}
\begin{equation}
\epsilon = \exp(\frac{\partial F(\nu_a,\nu_{be})}{\partial\nu_a}) \ \ .
\end{equation}
It is supposed that $\epsilon$ depends on $\nu_a$ rather weakly.
We suppose that $\epsilon$ is locally a constant value. This
supposition is many times weaker than the previous approximation.

Since
$\frac{\partial F(\nu_a,\nu_{be})}{\partial\nu_a}$ is small the
value of $\epsilon$ is close to $1$ and $1-\epsilon$ is very
small.
Then the value $1+\epsilon$ is close to $2$. Then the
relative deviation
of $\frac{\partial F(\nu_a,\nu_{be})}{\partial\nu_a}$
have no importance.

Then one can get
\begin{equation}
x = c y \ \ ,
\ \ \ \ N\exp(x) + A \frac{d^2N}{dx^2} + B \frac{dN}{dx} = 0
\ \
\end{equation}
with the known values of $A$, $B$.

After the transition to $\tilde{\psi} = \exp(x)$ one can get
\begin{equation}
A \tilde{\psi}^2 N'' + (A+B) \tilde{\psi} N' + \tilde{\psi} N = 0
\ \
\end{equation}
with the known solution
\begin{equation}
N = \tilde{\psi}^{-B/(2A)} Z_{B/A}(\frac{2}{\sqrt{A}}
\tilde{\psi}^{1/2}) \ \ ,
\end{equation}
where $Z_i$ is the cylinder function. One has to choose the
solution vanishing at $\infty$.

The known value of $N$ allows to
determine the total intensity of the embryo formation and the
integral can be taken analytically.

\subsection{Applicability of solution}

Our solution corresponds to the solution derived by H. Trinkaus in
\cite{Trink}. But this correspondence is only a formal one. Recall
the derivation by Trinkaus in \cite{Trink}. Trinkaus proposed the
linearization of the free energy $F$ ($G$ in terms of Trinkaus)
around $\hat{n}_2$ (this value is analogous to $\nu_{a0}$).

Now we shall analyze the possibility of linearization of $F$ in the
vicinity of $\nu_{a0}$. This linearization can be considered in the
global sense and in the local sense when linearization is done over
one coordinate while the other coordinate determines the values of
coefficients in this linearization.

Linearization in the global sense can not exist because the
second
derivative at the ridge and the
second derivative at the valley must have
different values. Only then the value of
$$
\Delta F(\nu_a) \equiv F(\nu_{a},\nu_{bx}) -  F(\nu_{a},\nu_{be})
$$
will be a real activation barrier.
The exponent of the last value is the leading term
in the expression for the flow.

Linearization in the local sense can not be valid also. It is
absolutely clear that the linearization over $\nu_b$ can not be
made because it is necessary to have a valley and a ridge for $F$
as a function of $\nu_b$. So, it can not be linearized. Another
possibility is to fulfill linearization over $\nu_a$ while
coefficients depend on $\nu_b$. The last possibility is the most
preferable one.

The careful analysis of the last possibility shows the
impossibility of linearization. Really, since the ridge in $\nu_a,
\nu_b$ scale is relatively close to the ridge in $\kappa$, $\xi$
scale one can see that the behavior of $F$ as a function of $\nu_a$
at $\nu_b$ slightly greater than $\nu_{bx}$ is the following one:
At first $F$ increases until the ridge in $\nu_a, \nu_b$ will be
attained. Later with increase of $\nu_b$ the value of $F$ will
decrease. This behavior is the direct consequence of the slope of
the channels of nucleation in $\nu_a, \nu_b$ plane. So, the
linearization is impossible.

The only possible variables,
in which the approximate local linearization
is valid are variables $\kappa, \xi$. One can see that there $F$
can be linearized far from the critical point
$$\partial F (\kappa, \xi) / \partial \kappa = 0 $$
at every $\xi$. The linearization is  made only
along $\kappa$. But these variables have not been even mentioned in
\cite{Trink}.

It has been already analytically shown that we are far from
the critical point.
Namely this allows the linearization in a local sense along $\kappa$.

The critical point which is the nearest to the origin of
coordinates is situated in the channel in $\nu_a, \nu_b$ picture.
This is the real saddle point. But since we are far from the
main saddle
point it means that we are far from every critical point.

Now we shall see that the linearization of the free energy in
$\kappa, \xi$ variables is possible. Really,
$$
\frac{\partial F(\kappa, \xi)}{ \partial \kappa}  =
 - b_g (\xi) + 2 \kappa^{-1/3} / 3
$$
The second derivative is
$$
\frac{\partial^2 F(\kappa, \xi)}{ \partial \kappa^2}  =
  - 2 \kappa^{-4/3} / 9
$$

The size of characteristic region in which the linearization is
necessary can be estimated as
$$
\Delta\kappa =(b_g (\xi) - 2 \kappa^{-1/3} / 3)^{-1}
$$
So, the necessary condition is
$$
|(- b_g (\xi) + 2 \kappa^{-1/3} / 3)^{-2} 2 \kappa^{-4/3} / 9 | \ll 1
$$
Since we are far from the critical point one can neglect the
compensation in $( - b_g (\xi) + 2 \kappa^{-1/3} / 3)$ and get
$$
|(2 \kappa^{-1/3} / 3)^{-2} 2 \kappa^{-4/3} / 9  | \ll 1
$$
or
$$
  \kappa^{-2/3}  \ll 1
$$
The last inequality is evident.

The last property is important for our needs.
We are interested in the linearization of the free energy of
the ridge and of the valley.
Really, the
particular case of the last
derivation is the possibility of linearization of $F$ along
the ridge and the valley in $\kappa, \xi$ scale, i.e. along
the channel and along the separation line.

The last step is to go from $\kappa, \xi$ picture to $\nu_a,
\nu_b$ picture.  We see that the slope of the valley and the ridge
in $\kappa, \xi$ picture
along $\kappa$ is very small. Since the slope is proportional to
$|\frac{\partial F(\kappa, \xi)}{ \partial \kappa}|$ it
can be seen from
$$
|\frac{\partial F(\kappa, \xi)}{ \partial \kappa} | =
| - b_g (\xi) + 2 \kappa^{-1/3} / 3 | \sim 2 \kappa^{-1/3} / 3  \ll 1
$$
So, the characteristic distance where the height of the valley,
 the height of the ridge and, thus, the height of the
 activation barrier (in fact it can be proven that there is no
 compensation) undergo the variation of one thermal unit    is
$$
D_1 = \kappa^{1/3}
$$
One can see that $D_1 \ll \kappa$ and it means that the relative
size of the transition region has to be small.

This slope has to compared with the characteristic halfwidth along
$\nu_b$ or
the characteristic size $D_1$
has to be compared with the
half-width along $\xi$ multiplied on $\kappa$.
We have
$$
D_2 = (\frac{\partial^2 F(\kappa, \xi) }{2\partial \xi^2})^{-1/2}
\kappa = (\frac{\partial^2 b_g( \xi) }{2\partial \xi^2})^{-1/2}
\kappa^{1/2} \sim \kappa^{1/2}
$$
We see that
$$
D_2 \ll D_1
$$

The slope at the boundary of halfwidth is
$$
\frac{\partial^2 F(\kappa, \xi) }{\partial \xi^2}  D_2 / \kappa
 \sim \kappa^{1/2}
$$
and it is rather essential.

We introduce the distance $D_3$ where the slope
$$
S_l =\frac{\partial^2 F(\kappa, \xi) }{\partial \xi^2} D_3
/ \kappa^2 \sim D_3 / \kappa
$$
has the order of the slope of the ridge
$\partial F / \partial \kappa \sim \kappa^{-1/3}$, i.e.
$\kappa^{-1/3}$.
Then we get
$$
D_3 \sim \kappa^{2/3}
$$
We see that the order of $D_3$ is the same as the order of $D_1$
and it is relatively small
$$
D_3 \ll \kappa
$$
 It means that the deviation of the
separation line
 in $\kappa, \xi$ scale from the ridge in $\nu_a, \nu_b$ scale
is relatively small.

Since $F_h$, $F_s$
allow linearization as functions of $\kappa$ or of $\nu_a$
 we come to a conclusion that the linearization of
 $F_e$, $F_r$, $\Delta
F$ (this value is a function of one variable) as a function of
$\kappa$ or of $\nu_a$ is quite possible.

\subsection{Simplified solution}

Since $\partial F(\nu_a, \nu_{be}) /\partial \nu_a \ll 1$ one can
put $\epsilon = 1$. Then $B=0$ and one come to the universal
solution
\begin{equation}
N \sim  Z_{0}(\frac{2}{\sqrt{A}}
\tilde{\psi}^{1/2})
\end{equation}
This is the universal function
$Z_0$ of the
variable
$$
\frac{2}{\sqrt{A}}
\exp(cx/2)
$$
Finally we get a universal solution.

\subsection{Discussion}

The multidimensional case is quite analogous to the two-dimensional
one. In the multidimensional nucleation one has to consider some
channel of nucleation.
One has to extract the set of fast variables $\{ \nu_b \}$ and the
set of slow variables $\{ \nu_a \}$.

For the set $\{ \nu_a  = fixed \}$
one can establish $J_{\{\nu_a\}}$ by
the consideration of the evolution in the set $\{\nu_b\}$.
It can be done  by the
standard methods from the previous sections.

After the calculation of $J_{\{\nu_a\}}$ one can define the
direction. It will be the quasi-integral on $\nu_a$. This defines
the first coordinate. The second coordinate is the direction of the
bottom of the valley in the cross section
$\{ \nu_b  = const \}$.
 The further consideration is absolutely analogous.

The new results formulated above are the following:

\begin{itemize}

\item
In the paper of Trinkaus \cite{Trink} only the differential case
was considered. The discrete case was not considered there.
 Really, the height of the pseudo-activation
barrier can change rather rapidly with increase of $\nu_a$.
This leads to the preference of discrete model.

As for the half-widths of the bottom of the channel and of the top
of the ridge in calculation of $J$ there are inequalities
which guarantee the possibility of the differential description.
Really, these half-widths increase like $\kappa^{1/2}$ (see the
standard estimates for the half-widths along the stable variables).
But if even these variables will be not so big nothing will be
changed because they variate slowly in comparison with the exponent
of the height of the pseudo-activation barrier. So, the mathematical
structure of the balance equation will be the same.

\item
Here the surface limited growth is considered while in
\cite{Trink} the diffusion limited growth was used. It seems
that because the transition occurs earlier than the saddle point
will be attained the embryos are small enough and the surface
limited growth is preferable.

\item
It is shown that the absence of the equilibrium distribution in the
pre-critical region is the driving force of the transition far from
the saddle point. This fact stresses once more the importance of
the formulation of the boundary conditions and outlines the paper
\cite{Mel2} where the boundary conditions were used for the
situation without hierarchy of kinetic coefficients.

\item
The hierarchy of the halfwidths of the near-critical region (more
accurate the near-saddle region) shows that the quasi-unary
nucleation in the square approximation in the neighborhood of the
saddle point is impossible. So, the change of approximation to a
linear one is absolutely necessary. Moreover, it is impossible to
see the transition of the Stauffer's solution to Trinkaus' one on
the analytic level of explicit formulas.

\end{itemize}

Beside the mentioned disadvantages of the differential approach one
can mention
the disadvantage
 connected with the position of the basic point
 $\nu_a^*$  for
decompositions of the height of the ridge and depth of
the valley. An
ordinary chosen point for such decompositions is
\begin{equation}\label{51a}
W_a^+
=J(\nu_a) / N(\nu_a)
\end{equation}
The presence of this point awakes the idea of
the Genuine Saddle Point
\cite{Li}.
It is reasonable to put the point of decomposition at
\begin{equation}
\label{51b}n(\nu_a^*) = n_{eq}(\nu_a^*)/2
\end{equation}
The shift between $\nu_a^*$ determined by (\ref{51a}) and
(\ref{51b}) will be called "the soft shift".

The greater is $|c|^{-1}$, the greater is the soft shift.
But the applicability of differential approach requires
$$
|c| \ll 1
$$
The last parameter
ordinary comes from two decompositions: one of the height
of the ridge\footnote{Take a cross section $\{ \nu_b  = const
\}$.}
$$
F_r(\nu_a) = F_r(\nu_{a0}) + k_r (\nu_a - \nu_{a0})
$$
with parameter $k_r$ and another of the depth of the
valley\footnote{Take a cross section $\{ \nu_b  = const
\}$.}
$$
F_e(\nu_a) = F_e(\nu_{a0}) + k_e (\nu_a - \nu_{a0})
$$
with parameter $k_e$.

Ordinary
$$ k_e  > 0 $$
(the opposite sign means that the saddle point is already behind)
$$ k_r <0 $$
(the opposite sign means that energetically it was more profitable
to cross the ridge earlier\footnote{Then the cross of the ridge can
not disturb the equilibrium distribution. So, the flow is known.}).
Then in
$$
J \simeq J_0 \exp( - k_r (\nu_a - \nu_{a0}) + k_e (\nu_a - \nu_{a0}) )
= J_0 \exp( c (\nu_a - \nu_{a0}))
$$
parameters $k_e$ and $k_r$ can not be compensated.
Ordinary both linear
approximations are necessary.

Then the condition $|c| \ll 1$ leads to
$$
|k_r| \ll 1
$$
$$
|k_e| \ll 1
$$

Under the last two inequalities one can see that $N$ becomes many
times less than the equilibrium value $N^{eq}$ much earlier than
$\nu_a = \nu_{a0}$ and the transition is actually over. So, the
point of decompositions has to shifted.

The shift of decompositions has to lead to the basic point situated
at the position characteristic for the relatively intensive flow.
One of the possible recipes is to choose the point $\nu_a^*$ of
decomposition according to
$$
N(\nu_a)
W_a^+ = \int_0^{\nu_a^*} J d \nu_a'
$$
The last condition can be approximately rewritten as
$$
W_a^+ = \frac{1}{|k_r| + |k_e| } \frac{1- J(\nu_a^*)}{N}
$$
One can start instead of $\nu_a = 0$ from infinity and get
a similar  estimate. Also it is reasonable to consider
$$
W_a^+ = \frac{1}{|k_r| + |k_e| } \frac{J(\nu_a^*)}{2N}
$$
as the point for decompositions.

Here naturally appears the length $\Delta$ of the region where the
transition occurs. It can be estimated as
$$
\Delta =\frac{1}{|k_r| + |k_e| }
$$
So, the soft shift can be greater than this region.

We continue to consider the problems of the differential approach.

\begin{itemize}

\item
Another problem is the smallness of $|k_e|$, $|k_r|$. Because of
the monotonous character of derivatives of the free energy along
channels and ridges it can be attained only near the saddle point.
But here the square approximation has to be used and the Stauffer's
solution will be the answer.

Certainly, if the value of $\nu_a$ is extremely big one can observe
small values of derivatives rather far from the saddle point. But,
although even here the discrete approach is preferable as
it will be shown later.

\end{itemize}

Now the simplified approximate method for continuous case will be
presented. In equation
$$
\frac{d^2 N}{d x^2} -
k_e\frac{d N}{d x} = N
\frac{\exp(F_r - F_e)}{\Delta \nu_e \Delta \nu_r}
\frac{W_{bx}^+}{W_a^+}
$$
one can put $k_e\frac{d N}{d x}$ to zero because of the smallness
of $|k_e|$.

Also because of the smallness of $|k_e|$, $|k_r|$ one can put
very approximately
$$\frac{\exp(F_r - F_e)}{\Delta \nu_e \Delta \nu_r}
\frac{W_{bx}^+}{W_a^+}
$$ to some
constant (let it be $I_0$). Then
$$
\frac{d^2 N}{d x^2} =
N I_0
$$
Solution of the last equation is evident
$$
N = A \exp( - \sqrt{I_0} x ) + B \exp( \sqrt{I_0} x )
$$

The requirement $N \rightarrow 0 $ at $x \rightarrow \infty$ leads
to
\begin{equation}
\label{53}N = A \exp( - \sqrt{I_0}  x )
\end{equation}

But this solution has a bad behavior at $x \rightarrow - \infty$.
So, in this region one has to use another approach. At $x
\rightarrow - \infty$ the flow is very small and $N$ is
approximately equal to the equilibrium value $N_{eq}$. Then
$$
\frac{d^2 N}{d x^2} -
k_e\frac{d N}{d x} = N_{eq}
\frac{\exp(F_r - F_e)}{\Delta \nu_e \Delta \nu_r}
\frac{W_{bx}^+}{W_a^+}
$$

Then approximately
$$
N = N_{eq} - \int J dx
$$
or
$$
N = N_{eq} - \int N_{eq} \frac{\exp(-(|k_e|+|k_r|) x )}{\Delta
\nu_r \Delta \nu_e} dx
\frac{W_{bx}^+}{W_a^+}
$$
With the evident approximation for the equilibrium value $N_{eq}$:
$$
N_{eq} = N_* \exp(-|k_e| x)
$$
with parameter $N_* = N_{eq} (x=0)$ one can get
$$
N = N_{eq} - N_* \int \frac{\exp(-|k_r| x )}{\Delta \nu_r \Delta
\nu_e} dx
\frac{W_{bx}^+}{W_a^+}
$$
Since one can approximately take $ \Delta \nu_r \Delta \nu_e $ as a
constant value there are no problems with integration. So,
$$
N = N_{eq} - N_* \frac{1- \exp(-|k_r| x )}{|k_r| \Delta \nu_r
\Delta \nu_e}
\frac{W_{bx}^+}{W_a^+}
$$

These two solutions (let it be $N_1$ and $N_2$) have to be stuck
together at the point where
$$
N_1 = N_2
$$

Another method can be formulated if we notice that
(\ref{53}) is valid namely locally because it was derived
with a supposition $I_0= const$. So, we have to go to the
local form by differentiation of (\ref{53}) which gives
$$
\frac{dN}{dx} = - N I_0
$$

This equation can be integrated with arbitrary $I_0$ which
leads to
$$
N \sim \exp(-\int I_0 (x') dx')
$$

When the evident known functional form
$$
I_0 \sim exp(cx)
$$
is taken, one can come to
$$
N \sim \exp(-\frac{I_{00}}{c} \exp(cx))
$$
with parameter $I_{00}$. Certainly, parameters $I_{00}$ and
$c$ can be considered here as the fitting parameters.

The functional form  announced above  resembles
$\Theta$-function with a soft transition from $1$ to $0$.
We shall call it as a soft $\Theta$-function and denote it
by
$$
S(x) = \exp(-\exp(x))
$$
This function can be  used as a brick in an ansatz
$$
Q = \sum A_i S(a_i(x-x_i))
$$
which can be very effectively used as an approximate
solution in all situations considered below in this paper.

\section{Interaction of valleys}

\subsection{Coordinates of valley in the $\nu_a, \nu_b$
coordinate system}

The coordinate of the valley is given by the condition
$$
\frac{\partial F (\nu_a, \nu_b) }{\partial \nu_b} = 0
$$
The straight differentiation of the free energy gives
$$
\frac{\partial F}{\partial \nu_b} =
\frac{d \gamma}{d \xi} \frac{\partial \xi}{\partial \nu_b}
S +
\gamma \frac{2}{3} S^{-1/2}
[v_b + \frac{\partial v_a }{\partial \xi}
\nu_a \frac{\partial \xi}{\partial  \nu_b}
+
\frac{\partial v_b }{\partial \xi}
\nu_b \frac{\partial \xi}{\partial  \nu_b}
]
$$
$$
 - b_b
-
[
\frac{\partial b_a }{\partial \xi}
\nu_a \frac{\partial \xi}{\partial  \nu_b}
+
\frac{\partial b_b }{\partial \xi}
\nu_b \frac{\partial \xi}{\partial  \nu_b}
]
$$
where $S$ is the surface square of the embryo. In simplest
approximation it can be written as
$$
S =  (v_a \nu_a + v_b \nu_b)^{2/3}
$$

The standard Gibbs-Duhem's equation looks like
$$
\nu_a d b_a + \nu_b d b_b = 0
$$
and leads to
$$
[
\frac{\partial b_a }{\partial \xi}
\nu_a \frac{\partial \xi}{\partial  \nu_b}
+
\frac{\partial b_b }{\partial \xi}
\nu_b \frac{\partial \xi}{\partial  \nu_b}
] = 0
$$
This brings the condition for the valley coordinate to
$$
\frac{\partial F}{\partial \nu_b} =
\frac{d \gamma}{d \xi} \frac{\partial \xi}{\partial \nu_b}
S +
\gamma \frac{2}{3} S^{-1/2}
[v_b + \frac{\partial v_a }{\partial \xi}
\nu_a \frac{\partial \xi}{\partial  \nu_b}
+
\frac{\partial v_b }{\partial \xi}
\nu_b \frac{\partial \xi}{\partial  \nu_b}
] - b_b
$$

But due to the surface enrichment the concentration differs from
$$
\xi = \frac{\nu_a}{\nu_a + \nu_b}
$$
and has to be
$$
\xi = \frac{\nu_a-  S \varrho_a}{\nu_a - S \varrho_a + \nu_b - S \varrho_b}
$$
Then the Giibs-Duhem's equation looks like
$$
S d \gamma + \nu_a d b_a + \nu_b d b_b = 0
$$
and leads to
$$
\frac{d \gamma}{d \xi} \frac{\partial \xi}{\partial \nu_b}
S + [
\frac{\partial b_a }{\partial \xi}
\nu_a \frac{\partial \xi}{\partial  \nu_b}
+
\frac{\partial b_b }{\partial \xi}
\nu_b \frac{\partial \xi}{\partial  \nu_b}
] = 0
$$
and
$$
\frac{\partial F}{\partial \nu_b} =
\gamma \frac{2}{3} S^{-1/2}
[v_b + \frac{\partial v_a }{\partial \xi}
\nu_a \frac{\partial \xi}{\partial  \nu_b}
+
\frac{\partial v_b }{\partial \xi}
\nu_b \frac{\partial \xi}{\partial  \nu_b}
] - b_b
$$

The careful analysis of the generalization of the
Gibbs-Duhem's
equation for the embryos shows that the terms
$$
\frac{\partial v_a }{\partial \xi}
\nu_a \frac{\partial \xi}{\partial  \nu_b}
+
\frac{\partial v_b }{\partial \xi}
\nu_b \frac{\partial \xi}{\partial  \nu_b}
$$
have to vanish
together with
$$
\frac{d \gamma}{d \xi} \frac{\partial \xi}{\partial \nu_b}
S + [
\frac{\partial b_a }{\partial \xi}
\nu_a \frac{\partial \xi}{\partial  \nu_b}
+
\frac{\partial b_b }{\partial \xi}
\nu_b \frac{\partial \xi}{\partial  \nu_b}
$$
 Really, the Kelvin's relation in the saddle point
requires that
\begin{equation} \label{u}
\frac{b_a}{v_a} = \frac{b_b}{v_b}
\end{equation}
The direct calculation with  a non zero value of the last terms gives
$$
\frac{b_b}{[v_b + \frac{\partial v_a }{\partial \xi}
\nu_a \frac{\partial \xi}{\partial  \nu_b}
+
\frac{\partial v_b }{\partial \xi}
\nu_b \frac{\partial \xi}{\partial  \nu_b}
]} =
\gamma (36 \pi)^{1/2}\frac{2}{3} S^{-1/2}
 =
\frac{b_a}{[v_a + \frac{\partial v_a }{\partial \xi}
\nu_a \frac{\partial \xi}{\partial  \nu_a}
+
\frac{\partial v_b }{\partial \xi}
\nu_b \frac{\partial \xi}{\partial  \nu_a}
]}
$$
and
one  can come to (\ref{u}) only if these terms vanish.

Generally speaking the Gibbs-Duhem's equation has to written in the
form
$$
\sum  (differentials \ of \  all \  intensive \
variables ) *  ( corresponding\ intensive \  variables )
 =  0
$$

Particularly
$$
S d \gamma + \nu_a d b_a + \nu_b d b_b + \nu_a d v_a + \nu_b d v_b
= 0
$$
Then
$$
\frac{d \gamma}{d \xi} \frac{\partial \xi}{\partial \nu_b}
S +
\gamma \frac{2}{3} S^{-1/2}
[ \frac{\partial v_a }{\partial \xi}
\nu_a \frac{\partial \xi}{\partial  \nu_b}
+
\frac{\partial v_b }{\partial \xi}
\nu_b \frac{\partial \xi}{\partial  \nu_b}
] - [
\frac{\partial b_a }{\partial \xi}
\nu_a \frac{\partial \xi}{\partial  \nu_b}
+
\frac{\partial b_b }{\partial \xi}
\nu_b \frac{\partial \xi}{\partial  \nu_b}
]
= 0
$$
and
$$
\frac{\partial F}{\partial \nu_b} =
\gamma \frac{2}{3} S^{-1/2}
v_b - b_b
$$

One can see that the concentration of valley satisfies the
condition
$$
\gamma \frac{2}{3 b_b (\xi)}
(v_a (\xi) + v_b (\xi^{-1} - 1) )^{-1/3} =\nu_a^{1/3}
$$
and it is not a constant value. Moreover, it is evident that
valleys in the $\nu_a, \nu_b$ system of coordinates do not
coincide with channels in $\kappa, \xi$ system of coordinates. They
coincide only in saddle points. The valleys in $\nu_a, \nu_b $
system can appear and disappear, their position in the absence of
hierarchy of kinetic coefficients means nothing.

\subsection{Asymptotics at $\nu_b \rightarrow \infty, \nu_a - fixed$ }

The necessary condition of applicability of solution of Trinkaus is
the limit
$$
\nu_b \rightarrow \infty , \  \nu_a = fixed  \ \ \ \ F \rightarrow - \infty
$$
The explicit calculation gives
$$
\nu_b \rightarrow \infty , \  \nu_a = fixed  \ \ \ \ F \rightarrow -
b_b \zeta_b
$$
So, it is necessary that $\zeta_b >0$. But the last condition is
not a necessary condition for nucleation in a gas mixture. The
necessary condition is the existence of concentration $\xi$ for
which the function $b_a \xi + b_b (1-\xi)$ is negative. So, there
exists a situation when there is no behavior necessary for
application of the Trinkaus' solution.

When $b_b>0$ one can see the asymptotic wing with a negative slope.
It will be called simply as "wing".

Otherwise there can be a situation when even with $b_b > 0$
nucleation can go from one valley to another (may be more deep)
valley and further no transition to $\nu_b \rightarrow \infty , \
\nu_a = fixed$ will take place because of the height of a new
(further) ridge.

In the case of purely supersaturated vapor of
components  the wings have to be
included in the general picture of relief of the free energy.

\subsection{Two valleys. Kinetic equation}

The previous consideration shows that the most ordinary situation
is the jump of embryo from one valley to the neighbor one. In one
valley (let it be called as the "source valley" and marked by the
subscript $-$) the embryos are in the pre-critical region (i.e.
$\kappa < \kappa_c$) and in the other valley (let it be called as
the "destination valley" and marked by the subscript $+$) the
embryos are in the post-critical region (i.e. $\kappa > \kappa_c$).
The transitions take place along lines $\nu_a = const$. Since the
increase of $\nu_b$ leads to the increase of $\kappa$ it is quite
possible.

The values of $\nu_b$ at the ridge will be marked as $\nu_{br}$.
The values of $\nu_b$ at the bottom of the source valley will be
marked as $\nu_{be-}$ and the values of $\nu_b$ at the bottom of
the destination valley will be marked as $\nu_{be+}$. All these
values are taken in the $\nu_a$, $\nu_b$ coordinate system.

Kinetic equations are rather transparent and look like
\begin{eqnarray}
\nonumber
\frac{d N_-}{dt} =
W^+_{as} (\nu_{a}-1,\nu_{be-})N_-( \nu_a -1 )
-W^+_{as}(\nu_{a},\nu_{be-}) N_-( \nu_a  )
\\
\\
+W^-_{as} (\nu_{a}+1,\nu_{be-})N_-( \nu_a +1 )
- W^-_{as}(\nu_{a},\nu_{be-}) N_-( \nu_a  )
- J_-  (\nu_a  )+ J_+ (\nu_a  )
\nonumber
\end{eqnarray}
\begin{eqnarray}
\nonumber
\frac{d N_+}{dt} =
W^+_{ad} (\nu_{a}-1,\nu_{be+})N_+( \nu_a -1 )
-W^+_{ad}(\nu_{a},\nu_{be+}) N_+( \nu_a  )
\\
\\
+W^-_{ad} (\nu_{a}+1,\nu_{be+})N_+( \nu_a +1 )
- W^-_{ad} (\nu_{a},\nu_{be+})N_+( \nu_a  )
- J_+ (\nu_a  ) + J_- (\nu_a  )
\nonumber
\end{eqnarray}
Here $N_-$ and $N_+$ are the numbers of embryos with given $\nu_a$
in a valley (in $\nu_a, \nu_b$ system of coordinates), $W_+$ and
$W_-$ are direct and inverse absorption coefficients, $J_-$ is the
flow from the source valley to the destination valley, $J_-$ is
the flow from the destination to the source valley (the inverse
flow).

We shall investigate the stationary solution.

One has to take into account that $W^+_a$ and $W_a^-$ are functions
of $\nu_b$. They are taken at $\nu_b$ equal to the values at the
bottom of valley. This can be done because of the relative
narrowness of valleys which goes from representation in $\kappa,
\xi$ variables.

\subsection{Two valleys. Direct and inverse flows}

The values of flows $J_-$ and $J_+$ are given by the standard
formulas
$$
J_-= N_- \frac{\exp(- F_r + F_{e-})}{\Delta_r \nu_b \Delta_{e-}
\nu_b} W^+_{bx}
$$
$$
J_+= N_+ \frac{\exp( - F_r + F_{e+})}{\Delta_r \nu_b \Delta_{e+}
\nu_b}W^+_{bx}
$$

Here $F_r$ is a free energy of the embryo at the ridge (in $\nu_a,
\nu_b$ coordinates), $F_{e-}$ is the free energy of the bottom of
the source valley in $\nu_a$, $\nu_b$ coordinate system, $F_{e+}$
is the free energy of the bottom of the destination valley in
$\nu_a$, $\nu_b$ coordinate system.

The value of $\Delta_r \nu_b$ is given by
$$
\Delta_r \nu_b = \sum_{\nu_b=\nu_{br1}}^{\nu_{br2}}
\exp(-F_r+F(\nu_a,\nu_b))
$$
Here $\nu_{br1}$ and $\nu_{br2}$ are chosen as roots of equation
$$
F(\nu_a, \nu_b) = ( 2 F_r + F_{e+} + F_{e-} )/4 $$ closest to
$\nu_{br}$ and
$$\nu_{br1} < \nu_{br} < \nu_{br2}$$

The value of $\Delta_{e-} \nu_b$ is given by
$$
\Delta_{e-} \nu_b = \sum_{\nu_b=\nu_{be-1}}^{\nu_{be-2}}
\exp(F_{e-}-F(\nu_a,\nu_b))
$$
Here $\nu_{be-1}$ and $\nu_{be-2}$ are chosen as roots of equation
$$F(\nu_a, \nu_b) = (   F_r +  F_{e-} )/2 $$ closest to
$\nu_{be-}$ and
$$\nu_{be-1} < \nu_{be-} < \nu_{be-2}$$

The value of $\Delta_{e+} \nu_b$ is given by
$$
\Delta_{e+} \nu_b = \sum_{\nu_b=\nu_{be+1}}^{\nu_{be+2}}
\exp(F_{e+}-F(\nu_a,\nu_b))
$$
Here $\nu_{be+1}$ and $\nu_{be+2}$ are chosen as roots of equation
$$F(\nu_a, \nu_b) = (   F_r +  F_{e+} )/2 $$ closest to
$\nu_{be-}$ and
$$\nu_{be+1} < \nu_{be+} < \nu_{be+2}$$

In continuous approximation one can get the following equations
$$
\Delta_r \nu_b = \int_{\nu_b=\nu_{br1}}^{\nu_{br2}}
\exp(-F_r+F(\nu_a,\nu_b))d\nu_b
$$
$$
\Delta_{e-} \nu_b = \int_{\nu_b=\nu_{be-1}}^{\nu_{be-2}}
\exp(F_{e-}-F(\nu_a,\nu_b))d\nu_b
$$
$$
\Delta_{e+} \nu_b = \int_{\nu_b=\nu_{be+1}}^{\nu_{be+2}}
\exp(F_{e+}-F(\nu_a,\nu_b)) d\nu_b
$$

One can prove that in frames of inequalities lying in the base of
capillary approximation the continuous approximation is valid. One
can also prove that in the absence of peculiarities in behavior of
the free energy it is possible in  the capillary approximation to use
the square approximation with infinite limits for calculation of
the mentioned values. This gives
$$
\Delta_{r} \nu_b = \sqrt{\pi}
( - \frac{1}{2} \frac{\partial^2 F(\nu_a,\nu_b)}{\partial
\nu_b^2}|_{\nu_b = \nu_{br}})^{-1/2}
$$
$$
\Delta_{e-} \nu_b = \sqrt{\pi}
( \frac{1}{2} \frac{\partial^2 F(\nu_a,\nu_b)}{\partial
\nu_b^2}|_{\nu_b = \nu_{be-}})^{-1/2}
$$
$$
\Delta_{e+} \nu_b = \sqrt{\pi}
( \frac{1}{2} \frac{\partial^2 F(\nu_a,\nu_b)}{\partial
\nu_b^2}|_{\nu_b = \nu_{be+}})^{-1/2}
$$

One can rewrite equations for $J_-$ and $J_+$ as following
$$
J_- = N_- I_-
$$
$$
J_+ = N_+ I_+
$$
where $I_+$ and $I_-$ are independent on $N_+$, $N_-$.

Already now one can fulfill the qualitative analysis of the kinetic
equations.

\subsection{Qualitative analysis of the kinetic
equations}

Consider the region of $\nu_a$ where $W^+_a
(\nu_{a},\nu_{be-})
\sim I_-$.
It is easy to see that at $\nu_a$ corresponding to the possible
transition from one valley to another
$$
W^+_a (\nu_{a},\nu_{be-}) < W^-_a (\nu_{a},\nu_{be-})
$$
(otherwise the saddle point in the source valley is already over)
$$
W^+_a (\nu_{a},\nu_{be+})
>
W^-_a (\nu_{a},\nu_{be+})
$$
(otherwise it will be necessary to overcome the saddle point in
the destination valley and it will cause the establishing of the
equilibrium distribution until the height of the saddle point;
moreover there is a straight way without barriers to the origin of
coordinates).

It means that
$$
F(\nu_a,\nu_{be-}) < F(\nu_a+1,\nu_{be-})
$$
$$
F(\nu_a,\nu_{be+}) > F(\nu_a+1,\nu_{be+})
$$

Moreover one can see that
$$
F(\nu_a,\nu_{br}) > F(\nu_a+1,\nu_{br})
$$
(otherwise it is more profitable to overcome the ridge earlier at
smaller $\nu_a$).

Practically in the main order
$$
\frac{I_-}{I_+} =
\exp(+ F_{e-} - F_{e+})
$$
The ratio $I_-/I_+$ governs the evolution of the process. One can
see two characteristic situations here
\begin{itemize}
\item
Situation
$$ I_- \gg I_+$$
Here one can see that the solution
of the previous section can be directly applied. The destination
valley do not affect the distribution in the source valley. So,
one can put $J_+=0$ and split the system of equations. Only the
first equation is essential and solution is really the solution in
the situation discussed above.
\item
Situation $$I_- \ll I_+$$ This situation has no analogs and has to
be considered separately.

\subsection{Situation $$I_- \ll I_+$$}

One can approximately put
$$
W^+_a (\nu_{a},\nu_{be-}) \simeq W^+_a (\nu_{a},\nu_{be+})
$$
This is taken only for simplicity.

Approximately,
the condition of the beginning of the jump of embryos, which
changes $N_-$ is the following
$$
I_- \geq W^+_a (\nu_{a},\nu_{be-})
$$
Then
$$
I_+ \gg W^+_a (\nu_{a},\nu_{be-})
$$
Then the second equation of the system becomes the following
$$
J_- = J_+
$$
and we have locally in a rough approximation
\begin{eqnarray}
\nonumber
\frac{d N_-}{dt} =
W^+_a (\nu_{a}-1,\nu_{be-})N_-( \nu_a -1 )
-W^+_a(\nu_{a},\nu_{be-}) N_-( \nu_a  )
\\
\\
+W^-_a (\nu_{a}+1,\nu_{be-})N_-( \nu_a +1 )
- W^-_a(\nu_{a},\nu_{be-}) N_-( \nu_a  )
\nonumber
\end{eqnarray}

Then
$$
\frac{N_-}{\Delta_{e+}\nu_b} = \frac{N_+}{\Delta_{e-}\nu_b}
\exp(-F_{e-}+F_{e+})
$$
and  approximately
$$
\frac{N_+}{N_-} =\exp(F_{e-}-F_{e+})
$$

The point where
$$I_- \approx W^+_a (\nu_{a},\nu_{be-}) $$
will be
marked as $\nu_a = y_0$. When $\nu_a$ increases one has
$$
I_- \gg W^+_a (\nu_{a},\nu_{be-}) \ \ \
$$
$$
I_+ \gg W^+_a (\nu_{a},\nu_{be-})
$$
This ensures the quasi-equilibrium and actually the common valley.
Later one attains $y_1$ where
$$
F_{e-}(y_1)  = F_{e+}(y_1)
$$
For $\nu_a > y_1$ one has
$$
I_- \gg I_+ \ \
$$
$$
J_- = J_+ \ \
$$
$$
N_+ \gg N_-
$$
It will be until $y_2$ defined by condition
$$
I_+(y_2) = W^+_a (\nu_{a}=y_2,\nu_{be-})
$$
(also the soft shift has to be added).
Later all remaining embryos from the source valley go into the
destination valley. But their total quantity is already rather
small. So, we need not to consider this process in details.

The main conclusion results in the appearance of the common valley
with a new free energy $F_0$. Here there is no connection with
the absence of excesses. This free energy can not be defined
separately from the width of the equilibrium distribution, only the
ratio
$$\exp(-F_0)/\Delta_{e0} \nu_b$$ can be determined. But namely
this ratio is the equilibrium distribution and in the expression
for the nucleation rate.

The last ratio can be determined from
$$
\frac{\exp(-F_0)}{\Delta_{e0} \nu_b} =
\frac{\exp(-F_{e-})}{\Delta_{e-} \nu_b} +
\frac{\exp(-F_{e+})}{\Delta_{e+} \nu_b}
$$
Very approximately one can say that
$$
\frac{\exp(-F_0)}{\Delta_{e0} \nu_b} =
\frac{\exp(-F_{e-})}{\Delta_{e-} \nu_b}
$$
when $F_{e-}<F_{e+}$ and
$$
\frac{\exp(-F_0)}{\Delta_{e0} \nu_b} =
\frac{\exp(-F_{e+})}{\Delta_{e+} \nu_b}
$$
when $F_{e-}>F_{e+}$.

\subsection{Intermediate situation}

\item
Intermediate situation is very rare because it can take place only
under the simultaneous realization of two equations
$$
I_- = I_+
$$
and
$$
I_- =W^+_a (\nu_{a},\nu_{be})
$$
(also the soft shift has to taken into account).
But this case in the only one where the interaction of valleys and
the exhaustion of the equilibrium distribution play simultaneously.

Solution of this situation is rather simple - it is necessary to
solve the system of several algebraic equations. At small
$$\nu_b<y_0$$ where
$$W^+_a (\nu_{a},\nu_{be-}) \gg I_-$$
 one has to use the
boundary condition
$$N_- = N_{-eq} \sim \exp(-F(\nu_a,
\nu_{be-})/\Delta_{e-} \nu_b$$
 $$N_+ \ll N_{+eq} \sim
 \exp(-F(\nu_a,
\nu_{be+})/\Delta_{e+} \nu_b$$
At big
$$\nu_b>y_0$$
where
$$W^+_a (\nu_{a},\nu_{be-}) \ll I_-$$ one has to use another
boundary condition
$$N_- \ll N_{-eq}$$
$$I_+ = 0$$
 if it will be necessary.
So, the task is to solve several simple algebraic equations.
Certainly, the discrete approach is preferable in
the computation.

\end{itemize}

To come to the continuous approximation one has to change the finite
differences for derivatives which
approximately leads to the following kinetic
equations
\begin{eqnarray}
\nonumber
\frac{\partial N_-}{\partial t} =
W^+_a (\nu_{a},\nu_{be-})
[k_{e-}\frac{\partial N_-}{\partial \nu_a} +
\frac{\partial^2 N_-}{\partial \nu_a^2}]
- J_-  (\nu_a  )+ J_+ (\nu_a  )
\end{eqnarray}
\begin{eqnarray}
\nonumber
\frac{\partial N_+}{\partial t} =
W^+_a (\nu_{a},\nu_{be+}) [\frac{\partial^2 N_+}{\partial \nu_a^2}
+ k_{e+}\frac{\partial N_+}{\partial \nu_a}]
- J_+  (\nu_a  )+ J_- (\nu_a  )
\end{eqnarray}

Here
$$
k_{e-} = - (1 - \exp(\partial F(\nu_a \nu_{be-})/\partial
\nu_a))
\approx
 \partial F(\nu_a \nu_{be-})/\partial \nu_a
$$
$$
k_{e+} = 1 - \exp(\partial F(\nu_a \nu_{be+})/\partial \nu_a)
\approx
- \partial F(\nu_a \nu_{be+})/\partial \nu_a
$$

Continuous approximation can not be widely spread but can be
applied only in rather specific situations. The reasons
are similar to those
described in analysis of the Trinkaus' solution. The proximity of
$d F_r / d \nu_b$ and both $d F_{e+} / d \nu_b$ and $d F_{e-} / d
\nu_b$ to zero means the proximity to the saddle point where the
linear approximation fails.

The simple approximate method is the iteration one - the values
$J_-$ and $J_+$ are calculated on the base of previous
approximations and they are treated as known functions. Initial
approximations are following:
\begin{itemize}
\item
when the source valley are deeper than the destination one, then
there is the quasi-equilibrium.
\item
when the destination valley are deeper than the source one, then
there is the Trinkaus' solution or the corresponding simplified
solution.
\end{itemize}

This method is very effective and leads to a rather precise
solution after one or two iterative steps.

It is necessary to stress here the effectiveness of the
method based on the ansatz with the soft Heavisaid's
functions.

The main result of this section
which was the goal  of the whole publication is the
radical change of the nucleation rate. The main goal is achieved -
the change of the nucleation rate in the orders of magnitudes is
shown. One can also see that the rate of nucleation
does not depend
on the free energy in saddle point but on the mutual position of
valleys and ridges and their relative heights. Certainly, the
problem to find the nucleation rate includes now the determination
of many characteristics and is more complex than in the theories
suggesting the recipes based on the value of the free energy in one
point. The theory presented here has to be used in order to get the
true value of the free energy. Now the problem is transformed in
the thermodynamic area - it is necessary to find the free energy of
the embryo formed in the mixture of vapors. This  problem is
complex enough to continue investigations of the binary and
multicomponent nucleation.

\section{Paths of transition}

Now one  can return to the general situation to see how the
real transition from the pre-critical region to the
post-critical region will occur.

The problem is to see where the real change of the channels will
take place. This problem will be solved here. So, here the
analysis will  be mainly qualitative. All details of transition
between channels will be a subject of a separate analysis.

\subsection{Approximate position of the valley}

To get the approximate position of valley and the ridge one
can act without surface excesses.

Consider the channel in coordinates $\nu_a, \xi$. Then
$$ \nu_a = \kappa  / p(\xi)
$$
where $p(\xi)$ is a known function
and
 $$ F =- B(\xi) p(\xi) \nu_a + p^{2/3} (\xi) \nu_a^{2/3}
$$

The coordinate of the valley
is given by condition
 $$
 \frac{\partial F }{\partial \xi} = 0
$$
or
$$ - B'(\xi) p(\xi) \nu_a - B(\xi) p'(\xi) \nu_a + \frac{2}{3}
p^{2/3} (\xi) \nu_a^{-1/3} p'(\xi)
=0
$$

 At the saddle point
 $$
- B(\xi) p'(\xi) \nu_a + \frac{2}{3}
p^{2/3} (\xi) \nu_a^{-1/3} p'(\xi)
=0
$$
and the saddle point of valley coincides with the saddle point of
the channel line, since
$$ B'(\xi) = 0$$

Asymptotically at $\nu_a \rightarrow \infty $ one can get
$$ - B'(\xi) p(\xi)   - B(\xi) p'(\xi)
=0
$$

One can see that the function $p$ is rather smooth while
$B$ is rather sharp. This  condition  is a definition of a "clear
channel".
Then one can neglect $B(\xi) p'(\xi)$ in comparison with
$B'(\xi) p(\xi)$. This leads to
$$    - B'(\xi) p(\xi)
=0
$$
and because of $p \neq 0 $ the last
equation coincides with the coordinate of the
channel.
So, we see that the valley is near the channel line for
every $\nu_a$.

To see the behavior at moderate $\nu_a$ near the critical values
one can note that $p'$ attains a moderate value. Really
$$
p' = \frac{\partial^2 \kappa }{\partial \nu_a^2}
\sim 1
$$
(we choose the space scale to have the volume for a molecule
in a liquid phase the order of $1$).
Then one has to take into account that
$$\kappa_c = 2/(3 max
\ B(\xi)) \gg 1 $$  requires $max \ B \ll 1
$
 Then the term
$ B(\xi) p'(\xi) \nu_a $ has a  small parameter. The term
$p^{-1/3} (\xi) \nu_a^{-1/3} p'(\xi)
$ has the same order as $ B(\xi) p'(\xi) \nu_a $ and, thus,
is small.
This reduces
the coordinate of a value to coordinate of a channel.

The same analysis can be done for every ridge. The general
approximate conclusion is that every separation line
corresponds to the ridge and their coordinates are similar.

At $\nu_a \rightarrow 0$ and $\kappa \rightarrow 0$ the
leading term is
$$
\frac{2}{3} p^{2/3} \nu_a^{2/3} p'
$$
which means that the valley  does not exist.  So, the valley
can not directly start at $\nu_a =0$ in continuous
approximation. Fortunately, ordinary this effect takes
place at $\nu_a$ less than $1$.

All above considerations are very approximate  and
they are
used only to see that qualitatively nothing is changes when
we consider valleys instead of channels.

Approximately speaking every channel corresponds to one
valley, their coordinates are rather similar.

Precisely speaking one can see many specific peculiarities,
for example, the appearance of valleys
without corresponding channels. But the probability of such
peculiarities is very low.
As a rule these valleys are not deep enough and can be
treated as negligible ones.

\subsection{Transition zones}

Consider the pair of valleys.

Every valley (index $v$)
can be considered as a source valley ($s$). Every
valley can be considered as a destination valley ($d$).
One can imagine many pairs of source and destination
valleys. Every pair has to be investigated.

At first we consider the situation when the channels are neighbor
ones.
The ridge is  the maximum of $F$ at the band $\nu_a =
const$ between the
concentration $\xi_s$ of a source valley
and the concentration $\xi_d$ of a  destination valley.

We define $F_v$ as the free energy at the valley, $F_r$ the
free energy at the ridge.

Now we shall make use from the approximate functional form
for $F_r$, $F_v$ established above
$$
F = const_1 \nu_{a}^{2/3} - const_2 \nu_a
$$

One can see the following facts
\begin{itemize}
\item
Every valley has only one critical $\nu_{avc}$ point determined by
$$
dF_v/d\nu_a = 0
$$
\item
One can define the pre-critical region
of the valley where $dF_v/d\nu_a >
0$ and post-critical region of valley where $dF_v/d\nu_a <
0$. There is only one pre-critical region with a size $\nu_a
< \nu_{avc}$ and a post-critical region where $\nu_a
> \nu_{avc}$.
\item
Every ridge has only one critical $\nu_{avc}$ point determined by
$$
dF_r/d\nu_a = 0
$$
\item
One can define the pre-critical region where $dF_r/d\nu_a >
0$ and post-critical region of ridge where $dF_r/d\nu_a <
0$. There is only one pre-critical region with a size $\nu_a
< \nu_{arc}$ and a post-critical region where $\nu_a
> \nu_{arc}$.
\end{itemize}

 The real effect on the nucleation rate
occurs when there is a transition from the pre-critical part of
the source valley to the post-critical part of a destination
valley. Transition from the post-critical part is useless because
the embryos can simply continue to grow, they already overcame the
barrier. So, there is no need to overcome
another one barrier and this case is out of our interest.

At first we suppose that in the whole pre-critical part of the
destination channel there is an equilibrium distribution. It
means that there is no further change of channels and the
destination channel will be the final destination channel. So,
there is only one cascade - only one change of the channel. We
shall call such processes as one-cascade processes.

One can choose components in such a way that the first component
is a rapid one.

Consider the regions where the probability
to change the channel is greater than to increase the value of
slow components  in the old channel.  This corresponds to condition
$$
W_{sl} \leq
W_1 Z_1 \exp(F_r - F_s)
$$
Here the kinetic coefficient $W_{sl}$ is the total kinetic coefficient
of all slow components, $W_1$ is the
kinetic coefficient of a rapid component and $Z_1$ is the corresponding Zel'dovich
factor for transition over the ridge.
The last inequality can be expressed in terms of the $F_r - F_v$ as
$$
F_r - F_v  \leq \ln(W_{sl}/(W_1 Z_1)) \equiv \Delta_t
$$
The rhs is a very slowly varying function.
Approximately  it is a constant.

Consider
$$
\Delta = F_r - F_v
$$
According to the approximate formulas the function
$\Delta (\nu_a)$ has the  second derivative
$$
\frac{d^2 \Delta}{d \nu_a^2} =
-[p^{2/3}(\xi_r) - p^{2/3}(\xi_v)] \frac{2}{9} \nu^{-4/3}_a
$$
which has a constant sign.

Thus, $\Delta$ has no more than maximum (it will be marked
by the index "m").

Certainly, the  condition
$F_r-F_v = \Delta_t$ depends on the scale of $\nu_a$. It
is necessary to choose the scale of $\nu_a$-axis to have
$$
d \Delta / d\nu_a  \sim 1
$$
at $\Delta \simeq \Delta_t$. Since $\Delta$ is not a too
sharp function of $\nu_a$, it is easy to do.
The condition $\Delta < \Delta_t$ can be
valid no more than in two zones:
one before $\nu_{am}$ another later $\nu_{am}$. Namely, in
intervals
$$
[ 0, \nu_{at-}], \ \ \ [\nu_{at+}, +\infty]
$$
the last condition is valid.

One can make the following notes:
\begin{itemize}
\item
The second interval $[\nu_{at+}, +\infty]$   can be absent when
$$
B(\xi_r) p(\xi_r) < B(\xi_s) p(\xi_s)
$$
\item
The first interval also can be effectively (not precise) absent
when $\nu_{at-}<1$ which occurs rather often.
\item
One can come to the situation when valleys are purely isolated.
\item
One has to keep in mind that the approximate formulas take
place only at big $\nu_a$.
\end{itemize}

The interval $[ 0, \nu_{at-}]$ will be called as
the "pre-transition zone", the interval $[\nu_{at+}, +\infty]$
- as the "post-transition" zone.

Consider the question about
 the mutual position of the destination and the source
channels.
The definition of $\kappa$ as  even without microscopic
corrections
$
(\sum \nu_i v_i) \gamma^{3/2}
$ contains $\gamma$ and $v_i$ and is a very complex function. But
in the majority of situations the increase of $\nu_a$
(other $\nu_i$ are fixed) causes the
increase of $\kappa$. We shall imply this property to
take place. This property will be referred  as the property
of
$\kappa$-convexity.
The line $\kappa = const$ as a function of $\nu_a$
is convex.

Certainly, in real systems there can be concave regions, where
the growth of $\nu_a$ leads to the change of concentration, the
partial volumes change, the surface tension change and the value
of $\kappa$ falls. But this situation is exclusive.

Under the property of convexity one can see that the transition
will be carried out only by addition of molecules of the first
component (ejection is not possible) and will go from
the left side to the
right side in $\nu_a, \nu_b$ plane.

The precise position of boundaries have to be defined with
surface excesses. Also a shift connected with a special
renormalization has to be taken into account.

\subsection{Nucleation conditions and supplying conditions}

Here we shall mark $\nu_a$ by $x$.

Conditions for the possibility of nucleation through the
post-transition zone are the following ones
\begin{itemize}
\item
Transition has to be effective, i.e.
there has to be a region in the post-critical region
in a destination valley $x>x_{dc}$ where $F_d(x)<F_{sc}$
Certainly this region looks like
$$[x_b, +\infty]$$
and the
beginning of this region has to be smaller than $x_{sc}$:
$$
x_b < x_{sc}
$$
\item
Transition has to be opened, i.e.
$$
x_{t+}<x_{sc}
$$
\end{itemize}

The beginning of transition will be at
$$
x_w = max \{ x_b, x_{t+} \}
$$

There are two possibilities at $x_w$:
\begin{itemize}
\item
The first possibility
$$F_d (x_w) > F_s (x_w)
$$
Here the common valley will be formed and the most effective
transition will be at $x_u$ defined as
$$F_d (x_u) = F_s (x_u)
$$
\item
The second possibility
$$F_d (x_w) < F_s (x_w)
$$
Here the transition from valley to valley occurs like a falling
from the high channel to the low channel. Solution looks like the
Trinkaus' one.
\end{itemize}

To see the real process of the channel transition it is necessary
to have corresponding conditions at the beginning of transition.
These conditions have to be the equilibrium conditions. It is
necessary that earlier in the valley there
would be no possibility to
escape from the valley. One has to analyze such possibility.

To
see the transition in the pre-transition zone
it is necessary that two conditions take
place
\begin{itemize}
\item
The first condition:
$$
F_d(x) < F_s(x_w)
$$
\item
The second condition:

Transition has to lead to the post-critical region in the
destination valley.
\end{itemize}

We are interested
to avoid such intensive transition which can destroy the
equilibrium conditions at $x_w$.

Since $F_d$ has to be at the post-critical region, it is a
decreasing function of $x$ and it is sufficient to check condition
at the boundary:
$$
F_d(x_{t-}) < F_s(x_w)
$$

In this
situation
the intensity of the valley transition in the pre-transition zone
is so big that there is no equilibrium condition for the
transition in the post-transition zone.

Since the transition in the pre-transition zone has to lead to the
post-critical zone then the peak of $F_d$
lies inside the transition through the pre-transition zone.
So, since
$$
max F_d > max F_s > F_s (x_w)$$ the transition occurs in a manner
of common channel and the real transition takes place at $x_p$ when
$$F_d (x_p) = F_s (x_p) $$ if $$x_p < x_{t-}$$

If at $$x=x_{t-}$$ we have $$F_d > F_s$$ the most  intensive
transition takes place at $x_{t-}$. This situation is more
probable than the precedent one.

What has to be done when the condition
$$
F_d(x_{t-}) > F_s(x_{t-})
$$
takes place?

Certainly, the transition can take place out of pre-transition and
post-transition zones but with a very low probability. To take
into account this possibility one has to add to $F_s$ the quantity
$F_r - F_s - \Delta_t$, i.e. to go from $F_s$ to $F_r- \Delta_t$.
This has to be done out of pre- and post-transition zones.

The point of transition will be near the root of equation
$$
F_d = F_r- \Delta_t
$$
Let it be at $x=x_y$.

This transition can not violate the equilibrium. So, the transition
in the post-transition zone is not destroyed and intensities
of this transition and transition in the post-transition zone
have
to be compared (added).

We shall call this transition as "the saturation transition".

Here the transition is going across the ridge into the valley. The
surface excesses can be taken into account very simply by noticing
that the forms of ridge and valley profiles remain the old ones
and only the shifts of profiles as a whole take place due to the
account of excesses.

\subsection{Details of the saturation transition}

Solution of the saturation transition is rather simple.
Consider at first the general situation.
Let $n_d(\nu_a) $ be the
embryos number density in a destination valley,
$n_s(\nu_a) $ be the
embryos number density in a source valley,
The evolution equation for the source valley looks like
$$
\frac{\partial n_s}{\partial t} =
 - \frac{\partial}{\partial \nu_a} W^+_{as} n^e_s(\nu_a)
 [ \frac{n_s(\nu_a)}{n^e_s(\nu_a)} -
 \frac{n_s(\nu_a+1)}{n^e_s(\nu_a+1)}] -
$$
$$
 n_s \frac{Z_s}{\Delta_s} \exp(-F_r+F_s)W_{bx}^+ +
n_d \frac{Z_d}{\Delta_d} \exp(-F_r+F_d)W_{bx}^+
$$

Here $W_a$ is kinetic coefficient, $n^e$ is the equilibrium
distribution, the flow
$$
n_s \frac{Z_s}{\Delta_s} \exp(-F_r+F_s)W_{bx}^+
$$
is the flow from the source valley to the destination valley
and
$$
n_d \frac{Z_d}{\Delta_d} \exp(-F_r+F_d)W_{bx}^+
$$
is the flow from the destination valley to the source valley.
The
value $Z$ is the Zeldovich' factor, $\Delta$ is the normalizing
factor.

Analogously one can write equation for the destination valley
$$
\frac{\partial n_d}{\partial t} =
 - \frac{\partial}{\partial \nu_a} W_{ad}^+ n^e_d(\nu_a)
 [ \frac{n_d(\nu_a)}{n^e_d(\nu_a)} -
 \frac{n_d(\nu_a+1)}{n^e_d(\nu_a+1)}] +
 $$
 $$
 n_s \frac{Z_s}{\Delta_s} \exp(-F_r+F_s)W_{bx}^+ -
n_d \frac{Z_d}{\Delta_d} \exp(-F_r+F_d)W_{bx}^+
$$

In continuous approximation
\begin{eqnarray}\label{p1}
\frac{\partial n_s}{\partial t} =
  W_{as}^+ [ \frac{\partial^2}{\partial \nu_a^2}  n_s(\nu_a)
+ \frac{\partial F_s }{\partial \nu_a}
\frac{\partial  }{\partial \nu_a} n_s(\nu_a)]
  \nonumber \\
  \\ \nonumber
 -
 n_s \frac{Z_s}{\Delta_s} \exp(-F_r+F_s)W_{bx}^+ +
n_d \frac{Z_d}{\Delta_d} \exp(-F_r+F_d)W_{bx}^+
\end{eqnarray}
for the source valley and
\begin{eqnarray}\label{p2}
\frac{\partial n_d}{\partial t} =
  W_{ad}^+ [ \frac{\partial^2}{\partial \nu_a^2}  n_d(\nu_a)
 + \frac{\partial F_d }{\partial \nu_a}
\frac{\partial  }{\partial \nu_a} n_d(\nu_a)]
 -
  \nonumber \\
  \\ \nonumber
 n_d \frac{Z_d}{\Delta_d} \exp(-F_r+F_d)W_{bx}^+ +
n_s \frac{Z_s}{\Delta_s} \exp(-F_r+F_s)W_{bx}^+
\end{eqnarray}
for the destination valley.

One can assume that
$$
\frac{\partial F_d }{\partial \nu_a}  = v_d
\ \ \ \ \ \
\frac{\partial F_s }{\partial \nu_a} = v_s
$$
are constants.
Also it can be assumed that the linear approximations
\begin{equation}
-F_r +F_d = A_d x+\tilde{C}_d
\end{equation}
\begin{equation}\label{73}
-F_r +F_s = A_s x+\tilde{C}_s
\end{equation}
for $x=\nu_a - \nu_{a0}$ are valid. Here $\nu_{a0}$ is some
parameter chosen as to belong to effective region of transition.

Then the stationary solutions will satisfy the system of equations
\begin{equation}\label{ss1}
\frac{\partial^2  }{\partial x^2} n_s + v_s
\frac{\partial  }{\partial x} n_s - C_s n_s \exp(A_s x)
+C_d n_d \exp(A_d x) =0
\end{equation}
\begin{equation}\label{ss2}
\frac{\partial^2  }{\partial x^2} n_d + v_d
\frac{\partial  }{\partial x} n_d - C_d n_d \exp(A_d x)
+C_s n_s \exp(A_s x)=0
\end{equation}
with
$$
C_s = \exp(\tilde{C_s}) Z_s W_{bx}^+/\Delta_s W_{as}
$$
$$
C_d = \exp(\tilde{C_d}) Z_d W_{bx}^+/\Delta_d W_{ad}
$$

In the second solution because the region in the destination
valley is the super-critical one it is possible to neglect
$\frac{\partial^2  }{\partial x^2} n_d $. Then the second equation
is the linear first order differential equation with a known
solution. Then after the substitution the first equation becomes
the closed the closed equation. Since solution of (\ref{ss2})
contains the integral then to get the differential equation it
will be necessary to differentiate (\ref{ss1}) one time and the
resulting differential equation will have the order 3. It can not
be solved at least in elementary functions.
So, it is necessary
to consider simplification based on classification
of transitions.

These are three types of transitions - the non-equilibrium falling
transition (first type), the equilibrium common valley transition
(second type),  the
equilibrium saturation transition (third type). For
different types we shall use different approximations.

For the first type it is possible to neglect
$$C_d n_d \exp(A_d x)$$
in the first equation. Then it becomes the closed equation
\begin{equation}\label{tr1}
\frac{\partial^2  }{\partial x^2} n_s - v_s
\frac{\partial  }{\partial x} n_s - C_s n_s \exp(A_s x)=0
\end{equation}
Then there is no necessity in validity of the linearization
(\ref{73}).

Solution  of the last equation
is presented above via cylindrical function.

The second equation is not necessary, but to complete the picture
one can write it in the form
\begin{equation}
 v_d
\frac{\partial  }{\partial x} n_d
+C_s n_s \exp(A_s x)=0
\end{equation}
with the solution
$$
n_d = - \int_{-\infty}^x C_s n_s \exp(A_s x')/ v_d dx'
$$

Certainly, the presentation of the solution via
the cylindrical function
is not convenient. It is more convenient to fulfill a block
transformation and then to solve the system of several algebraic
equations.
We shall formulate this procedure.
 Really, one can go from $x$ to $kx$ to have
 $$A_s k
\approx \ln \alpha $$
where the parameter  $\alpha\approx 1.5$.
Then with  an increase of $x$ by $1$ the intensity of
transition increases $1.5$ times. Then the equation
(\ref{tr1}) will be
\begin{equation}\label{tr2}
k^{-2} \frac{\partial^2  }{\partial x^2} n_s + v_s
k^{-1} \frac{\partial  }{\partial x} n_s - C_s n_s \alpha^x=0
\end{equation}

Now it is possible to consider the
interval $-2<x<2$ and to come back
to the initial discrete form of equation
\begin{equation}\label{tr3}
k^{-2} [ n_s(x+1)-2 n_s(x)+ n_s(x-1)] + v_s
k^{-1} [ n_s(x+1)- n_s(x-1)]/2 - C_s n_s(x) \alpha^x=0
\end{equation}
These coupled algebraic equations have to be written at
$x=-2,-1,0,1,2$.
At $x<-2$ one has to put $n_s$ to the equilibrium value. So, there
is a system of five coupled equations which can be easily solved.

One can continue analysis.
Every band has a separate physical meaning:
\begin{itemize}
\item
The band $x=2$ is the
starting band.
\item
At $x=-1$ one can use the smallness of the flow
$C_s n_s(x) \alpha^x$ and the smallness of the deviation of $n_s$
from the equilibrium value.
\item
The point $x=0$ is the point where $|d/dx[(n_s-n^e_s)/n^e_s]|$
attains maximum and, thus,
$$d^2/dx^2[(n_s-n^e_s)/n^e_s]=0$$
\item
At $x=1$ one can assume that $n_s$ is already small in comparison
with $n^e_s$
\item
The values at $x=2$ are the final values.
\end{itemize}
One can use these features at the characteristic zones to get
analytic solutions  and then the common solution will
be their combination.

These approximations allow to solve this equation
analytically by combination of the corresponding analytical band
solutions. But the resulting formulas will be very long to
use them
for calculations. To get concrete results it is more profitable to
solve algebraic equations, the precision is rather high while the
error is less than one tenth.

Certainly, one can directly solve the initial form of evolution
equation described earlier as the discrete model.

The analysis of the first type is completed.

For the transition of the second type it is possible to neglect
$$\frac{\partial^2  }{\partial x^2} n_s + v_s
\frac{\partial  }{\partial x} n_s$$
and
$$\frac{\partial^2  }{\partial x^2} n_d - v_d
\frac{\partial  }{\partial x} n_d $$
Then both equations will be reduced to
\begin{equation}
 - C_s n_s \exp(A_s x)
+C_d n_d \exp(A_d x)=0
\end{equation}
with the evident equilibrium solution as it was described
earlier. The point
$x=0$ is the saddle point, i.e. the point where approximately
$
F_s =F_d
$.

The third type of transition can be
described in  a following manner:
\begin{itemize}
\item
Equation (\ref{ss1}) leads to the fact that
$n_s = n_s^e$.
For $n_s^e$ one can take approximation
$$
n_s^e(x) = n_s^e(0) \exp(B_s x)
$$
where
$$
B_s = -  \frac{dF_s}{dx}|_{x=0}
$$
\item
Equation (\ref{ss2}) looks like
\begin{equation}
  v_d
\frac{\partial  }{\partial x} n_d - C_d n_d \exp(A_d x)
+C_s n_s^e \exp(A_s x)=0
\end{equation}
and can be easily solved since it is the first order linear
equation. The integral can be taken and the result will be
expressed via Whittaker or Kummer functions which can be reduced
to the Hypergeometric function.
\end{itemize}

The point $x=0$ has to be chosen as
$ \arg(max \frac{dn_d}{dx})$.

Since the result can be expressed only in the form of special
functions it is worth solving the discrete model. The method is
quite the same and one can come to
\begin{equation}
k^{-2}\frac{\partial^2  }{\partial x^2} n_d + k^{-1}v_d
\frac{\partial  }{\partial x} n_d - C_d n_d \exp(A_d k x)
+C_s n_s(0) \exp((B_s + A_s) k x)=0
\end{equation}
The value of $k$ has to be chosen to satisfy
$$ k \min(A_d, A_s+B_s) = \alpha $$
Then algebraic equations will be
\begin{eqnarray}
\nonumber
k^{-2}[ n_d(x+1) - 2 n_d(x) + n_d(x-1)] + k^{-1}v_d
[ n_d(x_1) -  n_d(x-1)]/2
\\
\\
\nonumber
- C_d n_d \exp(A_d k x)
+C_s n_s(0) \exp((B_s + A_s) k x)=0
\end{eqnarray}
and have to be written at $x=-2,-1,0,1,2$

Also it is necessary to mention the possibility to solve the
discrete model from the very beginning. The starting equation will
be
\begin{eqnarray}\label{ss22}
W_{ad}[ n_d(x-1)
- \frac{n_d^e(x-1)}{n^e_d(x)} n_d(x)
-n_d(x)
+ \frac{n_d^e(x)}{n^e_d(x+1)} n_d(x+1)]
\nonumber \\
\\ \nonumber
 - C_d n_d \exp(A_d x)
+C_s n_s^e \exp(A_s x)=0
\end{eqnarray}
These equations are coupled algebraic equations. The initial
condition is $n_d=0$ at $x \rightarrow - \infty$.

Our consideration has to be completed by equation on
parameters of transition.

The points of approximations $x=0$ form the  equations on
parameters of the process. The possible presence of the special
functions can be eliminated by  rational approximations for
special functions.
Then the parameters of th process will be determined by
the root of the
algebraic equation.

Now one can see the general picture of transition.
One can note that the complexity of the phase transition already
between two valleys is rather essential.
There are at least several possibilities to observe this
transition
\begin{itemize}
\item
Equilibrium transition in the pre-transition zone
\item
Equilibrium transition out of transition zone
\item
Equilibrium transition in the post-transition zone
\item
Falling transition in the post-transition zone
\end{itemize}
So, the unique approach
to get the rate of nucleation is impossible.

One has to stress that already
the equilibrium transitions can lead to
the absence of equilibrium in  valley with bigger $x$, and
$\kappa$. This effect has to be taken into account to diminish the
intensity of transition in the post-transition zone.

Here it becomes clear that the flow can split and merge.
Beside these effects
one can see the rapid change of the leading manner of
the supercritical embryo formation. This is caused only by kinetic
coefficients and, thus, it is reasonable to speak about "the
kinetic  rupture in the rate of nucleation".

One has to stress that in the saturation transition there
is no difference whether the transition occurs in the open
or in the forbidden zone. Really, $F_r - F_s $ can be
greater than $\Delta_t$:
$$F_r - F_s  >\Delta_t $$
and the transition will take place. The only conditions is
\begin{itemize}
 \item
$$
F_r (\nu_a) - \Delta_t  < F_{mc}
$$
where $F_{mc}$ is the value at the saddle point with a
minimal height
and
\item
$$F_d(\nu_a) < F_{sc}
$$
for some $\nu_a>\nu_{acd}$
\end{itemize}

Rigorously speaking the same consideration can take place
for transition of other types.

Then one can come to the situation when both the equilibrium
common channel transition and the falling transition can
take place. When at some $\nu_a > \nu_{acd}$
$$
F_d (\nu_a) = F_s(\nu_a) < F_{mc}$$
we have to examine $F_r$.

If
$$
F_r - \Delta_t < F_{mc}$$ we see that the intensity of the
common channel transition is greater than the intensity of
transition through the saddle point.

Since $F_d$ for $\nu_a > \nu_{acd}$ is a decreasing
function this intensity is also greater than the intensity
of a saturation transition. Then we have to compare it with
the intensity of a falling transition.

If
$$
F_r - \Delta_t < F_s
$$
we have the common valley transition  with intensity
greater than the further falling transition.

If
$$
F_r - \Delta_t > F_s
$$
then the further falling transition
will occur with intensity greater than the intensity
of the common valley transition.

All this is done without account of the soft shift. To take
this shift into account in a rough approximation  it is
necessary to add to $\Delta_t$ the quantity
$$
\ln[\frac{d(F_r - F_s)}{d\nu_a}]^{-1}
$$

\subsection{Other peculiarities of transition}

The property of the
$\kappa$-convexity is very important in the context of the
current analysis because it forbids the possibility to reach
the pre-critical region  after
the transition through the post-transition zone. Otherwise there
would be a source of embryos in some region of the destination
valley. The property of the
$\kappa$-convexity forbid the localization of the flow.

Such a localization of the transition flow can be seen in a
multi-valley transition. Consider the situation when
 there is an intermediate
valley (index $i$) and, thus, there are
two ridges - one between the source valley and
the intermediate valley (index $rs$), another
between the intermediate valley and the destination valley (index
$rd$).
Suppose that for intermediate valley $$F_{rs}-F_i
<\Delta_t$$
 $$F_{rd}-F_i <\Delta_t$$ Then one can speak about one effective
 ridge with a height
 $$
 F_{rr} = max (F_{rs}, F_{rd})
 $$
 Then the property of the ridge convexity disappears and the
localization of the transition flow can be seen.
One can speak about the
\begin{itemize}
\item
Injection at the finite zone into the valley.
\end{itemize}

Earlier we consider only two components in the mixture. So, the
inverse transition has to be the  backward one. But in the
three component mixture one can imagine the curved
transitions - at first transition
the concentration of the first
component increases, at the second transition the concentration of
the second component increases.
However, it is necessary to have at least two rapid components in
the mixture.
In some special cases it is possible to return to the same valley
but in another place of this valley - may be it is possible to
jump from the pre-critical zone to the post-critical one, may be
it is possible to make one loop of a spiral.
Here the picture will be
very picturesque.
However, it would be very nice to see the concrete examples of
such nucleating systems.

Here we do not consider the transitions from the
post-critical zone of one valley to the post-critical zone
of the other valley because this transition can not change
the rate of nucleation.

Despite the transition will have now a very complex form the
elementary bricks remain the same:
\begin{itemize}
\item
the equilibrium common-valley
transition
\item the equilibrium saturation transition
\item
 the
non-equilibrium falling transition
\end{itemize}

The possibility to reach the rather transparent
classification is based on the following simple approximate
structure of every channel/separation line/valley/ridge:
\begin{itemize}
\item
Every channel/separation line/valley/ridge has a
pre-critical zone which is directly (without hills) connected
with the origin, post-critical region with the irreversible
growth (until infinity) and a small near-critical growth.
\end{itemize}
The last property takes place for every channel, separation line,
valley and ridge.

According to the  last analysis the multi-cascade
transitions are not effective. Really, the cascade can lead
to the post-critical region or to the pre-critical region.
When it leads to the post-critical region it will be the
last cascade. If it leads to the pre-critical region there
is a smooth increasing path along the valley and this path
will have at least the same intensity. So, the transition
across the ridge is not effective here. As the result we
see that there is only one main cascade in the
multi-cascade transition.

Here we imply that one cascade can be the saturation
transition, the falling transition or the common valley
transition. Actually, the saturation transition is also the
common valley transformation because here there exists a
common valley. Then we shall speak here about the generalized
common
valley. Then there is the generalized
common
valley and the falling transition.

Certainly, the multi-common
valley can be such a cascade. In this cascade at some may be
finite zone several valleys are treated as one common
channel. It is also possible that the set of common
channels with the given channel can change - at some zone
there is one set, at another zone there is another set.
But in this common valley under the $\kappa$-convexity
there will be only one leading pair of the neighbor channels.

As the result we see that in the binary case
there is only one leading cascade which is the falling
transition or the generalized common valley.

We have examined only the stationary solutions. The
relaxation of the distribution $n$
to the stationary solution can be easily studied
since in all situations the stationary solution $n_{st}$
is known. Then one can linearize equation
on $n-n_{st}$ and get
$$
\frac{\partial n}{\partial t} = L n
$$
where $L$ is a differential operator (or in finite
differences) on $\nu_i$.
Then one can get
the relaxation time as the minimal
eigenvalue of the linear operator $L$
in the evolution equation by the iteration
procedures
$$Trace(L), \ \ Trace(L^2),  \ \ Trace(L^3), \ \ etc.$$
of the standard
numerical methods.

In reality all operators in the rhs of kinetic equations
are reduced to the square approximations. Then the
eigenvalues and
eigenfunctions are known. Eigenfunctions are the Hermite's
polynomials or the Generalized error-functions.

\section*{Main results}

One can see that the problem to get even the stationary rate of
nucleation is rather complex. Below we present the sequence of
actions to fulfill this task:
\begin{enumerate}
\item
We determine all channels and find the channel with a minimal
activation barrier. Determine its height $F_{cm}$
\item
We determine the rate $W_a/W_b$. Choose components to have $W_a <
W_b$ If $W_b/W_a < exp(1)$ there will be a Stauffer's solution
with $F_{cm}$. If there is an opposite situation one has to
continue consideration.

Suppose that we have the binary case and the $\kappa$-convexity.
The last property is rather ordinary but it simplifies the
consideration.  Then the procedure will be the following
\begin{enumerate}
\item
Instead of channels determine the valleys. D
We determine also all
ridges. We enumerate valleys  to have $\xi_i < \xi_j$ for $i<j$.
We enumerate  ridges to have $\xi_i < \xi_j$ for $i<j$. For every
neighbor valleys  we determine the source valley $i$ and the
destination valley $i+1$. Below we shall consider the one-cascade
transition.
\item
We determine the possibility of the saturation transition: there is
$\nu_a$ satisfying conditions:
$$
\nu_a < \nu_{acs}
$$
$$
\nu_a < \nu_{acr}
$$
$$
F_r(\nu_a) - \Delta_t < F_{cm}
$$
If these conditions are satisfied  we determine the point of the
saturated transition $\nu_a^*$ by equation
$$
F_d(\nu_a^*) = F_r(\nu_a^*) - \Delta_t
$$
This gives $$
F_* = F_d(\nu_a^*)
$$
\item
We determine
the possibility of the common valley   transition: there
exists $\nu_a$ with properties
$$
\nu_a < \nu_{acs}
$$
$$
\nu_a > \nu_{acd}
$$
$$
F_s(\nu_a) = F_d(\nu_a) < F_r (\nu_a) - \Delta_t
$$
The last condition determines only one point $\nu_a^{**}$ of the
common valley transition with a maximal intensity. This value will
be the saddle point of the unified valley. Here we determine
$$
F_{**} = F_s(\nu_a)
$$
If the equilibrium valley transition takes place there is no need
to consider the falling transition. If it does not exist then we
consider the falling transition.
\item
The falling transition takes place when there is $\nu_a$
satisfying conditions:
$$
\nu_a < \nu_{acs}
$$
$$
\nu_a > \nu_{acd}
$$
$$
F_d(\nu_a) < F_s(\nu_a)
$$
$$
F_r(\nu_a) - F_s(\nu_a) \leq \Delta_t
$$
Conditions
$$
 F_r(\nu_a) - F_s(\nu_a) = \Delta_t
 \ \
\ \ F_d(\nu_a) < F_s(\nu_a)
$$
determine the point of transition $\nu_a^{***}$ and the free
energy
$$
F_{***} = F_{s}(\nu_a^{***})
$$
\item
To see what transition is more profitable one has to compare
$F_{cs}$,  $F_*$, $F_{**}$ and $F_{***}$ and to choose the minimal
value
$$
F_{ch} = min(F_{cs},  F_*, F_{**} F_{***})
$$
This will be the free energy corresponding to this valley as the
source valley. Then one has to take the minimum over all valleys
and to determine the free energy of nucleation $F_f$. Then the
rate of nucleation is rather approximately given by
$$
J = \exp(F_f) Z W_a
$$
where the Zeldovich' factor $Z$ contains the normalizing factor of
the equilibrium distribution.
\item
If there are two approximately equal minimal values of free
energies between $F_{cs}$,  $F_*$, $F_{**}$ and $F_{***}$ then one
has to add the quantity $\ln(d(F_r - F_s)/d\nu)$ to the free
energy of the falling transition. Certainly, the last quantity
shifts the point of transition but approximately one can take it
at the old point.
\end{enumerate}

\end{enumerate}

The analysis presented above gave the following new results
\begin{itemize}
\item
The free energy of the embryo is found including the surface
excesses and the clear interpretation of the generalized chemical
potential is given. The variables giving the simple form of the
free energy is found and their connection with the initial natural
variables is shown (section 1). The similarity of the form of
the near-critical
energy  to the situation without surface excesses is shown
(section 1). The correction order of the Renninger-Wilemski's effect
is shown (section 1).
\item
The hierarchy in the near-critical region is shown (section 2).
\item
The impossibility of the essential difference between the Reiss'
formula ad the Stauffer's formula in capillary approximation is
shown (section 3).
\item
The possibility to change valleys during the evolution
 is shown. The discrete analog of Trinkaus' solution is
presented  and investigated (section 4).
\item
The possibility to have one united valley instead of several
initial ones is shown. It is shown that the height of the
effective
activation barrier is changed in comparison with the heights of
barriers in the initial channels. Thus, the rate of nucleation will
be radically changed (section 5)
\item
The possibility to change the valley and to reach the
post-critical zone already from the pre-critical transition
is shown (section 6). This form a  special type of the
equilibrium saturation transition. This transition also
leads to a new special value of effective height of
activation barrier.
\item
All possible transition are classified. It is shown that
the tree mentioned types  cover the variety of possible
transitions.
\end{itemize}

Here only the main new results are outlined. An application of
the presented theory to the concrete binary and multicomponent
systems will form the
subject of a separate publication.

\end{document}